\documentclass[final,3p,times]{elsarticle}

\usepackage{amsmath}
\usepackage{amssymb}
\usepackage{siunitx}

\newcounter{bla}

\journal{Computer Physics Communications}
\usepackage{graphicx}
\usepackage{bm}

\usepackage{hyperref}
\usepackage[capitalize]{cleveref}
\usepackage{csquotes}
\usepackage{dsfont}
\usepackage{booktabs}
\usepackage{listings}
\usepackage{xcolor}
\usepackage{makecell}

\newcommand*{\dd}{\ensuremath{\mathrm d}}
\newcommand*{\ee}{\ensuremath{\mathrm e}}
\newcommand*{\ii}{\ensuremath{\mathrm i}}
\newcommand*{\ex}{\ensuremath{\bm{\hat{x}}}}
\newcommand*{\ey}{\ensuremath{\bm{\hat{y}}}}
\newcommand*{\ez}{\ensuremath{\bm{\hat{z}}}}
\newcommand*{\erho}{\ensuremath{\bm{\hat{\rho}}}}
\newcommand*{\ephi}{\ensuremath{\bm{\hat{\phi}}}}
\newcommand*{\er}{\ensuremath{\bm{\hat{r}}}}
\newcommand*{\etheta}{\ensuremath{\bm{\hat{\theta}}}}
\newcommand{\treams}{\emph{treams}}

\renewcommand*{\phi}{\varphi}

\let\Re\relax
\DeclareMathOperator{\Re}{Re}

\DeclareMathOperator{\sign}{sign}

\crefname{lstlisting}{Listing}{Listings}
\Crefname{lstlisting}{Listing}{Listings}

\begin{document}

\begin{frontmatter}



\title{\treams{} -- A T-matrix scattering code for nanophotonic computations}


\author[a]{Dominik Beutel\corref{author}}%
\date{\today}
\author[b]{Ivan Fernandez-Corbaton}
\author[a,b]{Carsten Rockstuhl}
\address[a]{Institute of Theoretical Solid State Physics, Karlsruhe Institute of Technology (KIT), 76131 Karlsruhe, Germany}
\address[b]{Institute of Nanotechnology, Karlsruhe Institute of Technology (KIT), 76344 Eggenstein-Leopoldshafen, Germany}

\cortext[author]{Corresponding author.\\\textit{E-mail address:} dominik.beutel@kit.edu}

\begin{abstract}
We report the publication of \treams{}, a new software for electromagnetic scattering computations based on the T-matrix method. Besides conventional T-matrix calculations for individual scatterers and finite clusters of particles, a unique feature of the code is its full support for periodic boundaries in one, two, and all three spatial dimensions. We use highly efficient and quickly converging lattice summation techniques based on the Ewald method to evaluate the arising lattice sums in these cases. In addition to the common use of vector spherical waves as a basis set for the T-matrix, vector cylindrical waves are also implemented. To describe stratified media, vector plane waves are used with an S-matrix description of the electromagnetic scattering. All basis sets and the associated methods can be used together with chiral constitutive relations.

This contribution outlines the basic methods implemented and the program structure. Two interfaces to the implemented functionality are available: a flexible and fast low-level interface and a high-level interface for added convenience and plausibility checks. We conclude with two examples: a demonstration of the field calculation in various lattices and the explorations of quasi-bound states in the continuum. The presented code was already used in calculations for various physical systems: from the mode properties of molecular arrays in cavities to analytical models for metasurfaces and from moir\'{e} lattices to the homogenization of artificial photonic materials. With the publication of \treams{} and the associated documentation, we hope to empower more scientists to make an efficient, fast, and precise exploration of nanophotonic systems that can be described in the broader framework of scattering theory.
The full open-source code can be obtained from \url{https://github.com/tfp-photonics/treams} and the latest documentation is available online at \url{https://tfp-photonics.github.io/treams}.
\end{abstract}


\end{frontmatter}



{\bf PROGRAM SUMMARY}

\begin{small}
\noindent
{\em Program Title:} treams\\
{\em CPC Library link to program files:} (to be added by Technical Editor) \\
{\em Developer's repository link:} https://github.com/tfp-photonics/treams \\
{\em Code Ocean capsule:} (to be added by Technical Editor)\\
{\em Licensing provisions:} MIT  \\
{\em Programming language:} Python and Cython \\
{\em Nature of problem:}
Simulating electromagnetic scattering in periodic nanophotonic structures, when different length scales are present or for large parameter sweeps, requires specialized tools that can solve Maxwell's equations more efficiently than general-purpose solvers. A possible tool for these computations is the T-matrix method, which needs to be amended by suitable lattice summation schemes in the presence of periodic boundary conditions.
\\
{\em Solution method:}
The properties of the individual scatterers are described by T-matrices, such that the interaction can be solved analytically. The slowly converging lattice sums that appear in the presence of periodic boundary conditions are computed by converting them to two quickly converging series using Ewald's method. Depending on the lattice dimensions and the geometry of the scatterers, vector spherical, cylindrical, or plane waves are used. Using modes of well-defined helicity enables the straightforward inclusion of chiral material parameters.
\end{small}

\section{Introduction\label{sec:intro}}
Advances in fabrication technology for nanophotonic devices open the avenue for novel designs and applications. This progress in manufacturing needs to be accompanied by novel simulation techniques that can outperform general-purpose Maxwell solvers, e.g., when hugely different length scales or large parameter sweeps are involved. The T-matrix (transition matrix) method~\cite{waterman1965} has proven to be an exceptionally useful tool to tackle such scattering scenarios. Various programs to solve multi-scattering scenarios in different conditions have been developed~\cite{stefanou2000,mackowski2011,somerville2016,markkanen2017,egel2017a,necada2021,schebarchov2022}, and a continuously expanding list of software is available online~\cite{wriedt2008a,hellmers2009}.

While each of these codes has its advantage and unique capabilities, there is a lack of support for arbitrary periodic boundary conditions, especially when complex unit cells, i.e., unit cells containing multiple particles, are involved. Furthermore, including modes with well-defined helicity required to treat chiral constitutive relations is usually missing.
Therefore, we present \treams{} a program to solve electromagnetic scattering problems. It combines the highly efficient description of scatterers by the T-matrix formalism with quickly convergent lattice sums to handle their possible periodic arrangement~\cite{linton2010,popov2014,beutel2023}. Furthermore, stratified structures can be included with an S-matrix (scattering matrix) description. Besides the conventional T-matrices that build on a vector spherical wave (SW) basis, vector cylindrical wave (CW) T-matrices~\cite{huang2019} are also part of \treams. The combination with the lattice sums for arbitrary one-dimensional (1D), two-dimensional (2D), and three-dimensional (3D) arrays of the translation coefficients for these underlying basis sets allows an efficient description of all types of arrays. Moreover, typical T-matrix operations, such as coupling within finite clusters, rotations, or translations, are implemented to support the simulation of various scattering scenarios.
The program is written using Python, while most calculation-heavy functions are accelerated using Cython and, thus, run as compiled C code.

The rest of this manuscript is structured as follows. We first define the basis sets of well-defined parity and helicity in \cref{sec:scattering:waves}: SWs, CWs, and vector plane waves (PWs). These basis sets and their relations are important for most calculations in \treams.
Next, we will briefly overview the T-matrix method in \cref{sec:scattering:tmatrix}. We particularly exploit a generalized concept of the usual T-matrix definition based on SWs to also include CWs. These two basis sets complement each other on the type of object they can describe. Then, we outline the S-matrix method used for stacked planar structures in \cref{sec:scattering:smatrix}, which is used for the PW description. We conclude the first part with an overview of additional operators implemented in \cref{sec:scattering:ops}.

\Cref{sec:treams:structure} then gives an overview of how the main functions, methods, and operators are included in the framework of \treams. The section also describes other added functionality, such as importing and exporting T-matrix data in a file based on the HDF5 format~\cite{folk2011}. Moreover, the operators, such as translations or rotations, are explained more in-depth from the programming and usage perspective in \cref{sec:treams:ops}. These operators are working closely together with the implemented classes for general \enquote{physics-aware} arrays, i.e., \emph{numpy} arrays with added information about physical quantities such as the wave number, and the more specialized data structures for T-matrices and S-matrices. Those classes are the focus of \cref{sec:treams:arrays}.
Finally, we give examples of the capabilities of \treams. We start with technical examples on 1D and 2D arrays, highlighting how the different lattice sums and basis sets work together and complement each other in \cref{sec:examples:fields}. Finally, we show a physics-inspired example of calculating quasi-bound states in the continuum (quasi-BICs) in \cref{sec:examples:bic}.

\section{Electromagnetic scattering in chiral media\label{sec:scattering}}

This section summarizes the underlying electromagnetic scattering theory, which is the foundation of the program \treams. It describes SW, CW, and PW analytical solutions to the scattering of electromagnetic waves in chiral media. Using the first two of these basis sets allows the use of the T-matrix method, which is introduced for multi-scattering scenarios with and without periodic boundary conditions. We formulate these scattering scenarios in a global and a local T-matrix basis~\cite{suryadharma2017a}. Together with the lattice sums for the periodic boundary conditions, this allows us to describe complex unit cells, namely unit cells with multiple different particles. For PWs, we use the S-matrix method. Finally, we will shortly discuss additional operators, e.g., for rotations, the change of basis set, and other common transformations in electromagnetic scattering calculations.

\subsection{Vector spherical, cylindrical, and plane waves}\label{sec:scattering:waves}

\treams{} solves Maxwell's equations without free charges and currents. We assume monochromatic waves of vacuum wave number $k_0 = \tfrac{\omega}{c}$ where the time domain fields are defined as $\Re(\bm E(\bm r, k_0) \ee^{-\ii c k_0 t})$ with the angular frequency $\omega$ and speed of light in vacuum $c$. The assumption of monochromatic waves is justified due to the later choice of linear constitutive relations. With this ansatz, Maxwell's equations are
\begin{subequations}
    \begin{align}\label{eq:maxwell:curl}
        \nabla \times
        \begin{pmatrix}
            \bm E(\bm r, k_0) \\
            Z_0 \bm H(\bm r, k_0)
        \end{pmatrix}
        &=
        \mathrm{i} k_0
        \begin{pmatrix}
            0 & \mathds{1} \\
            -\mathds{1} & 0
        \end{pmatrix}
        \begin{pmatrix}
            \frac{1}{\epsilon_0} \bm D(\bm r, k_0) \\
            c \bm B(\bm r, k_0)
        \end{pmatrix}
        \\ \label{eq:maxwell:div}
        \nabla \cdot
        \begin{pmatrix}
            \frac{1}{\epsilon_0} \bm D(\bm r, k_0) \\
            c \bm B(\bm r, k_0)
        \end{pmatrix}
        &= 0\,,
    \end{align}
\end{subequations}
where all fields are normalized such that they have the same units as the electric field $\bm E(\bm r, k_0)$: the magnetic field $\bm H(\bm r, k_0)$ is multiplied by the free space impedance $Z_0$, the displacement field $\bm D(\bm r, k_0)$ is divided by the vacuum permittivity $\epsilon_0$, and the magnetic flux density $\bm B(\bm r, k_0)$ is multiplied by the speed of light. Describing the frequency by the wave number is convenient for the implementation later since no unit of frequency is used explicitly, and hence there is no need to keep track of it. Instead, all quantities, such as positions or wave numbers, are implicitly assumed to have the same unit of length or inverse length.

Maxwell's equations in this form need to be supplemented by constitutive relations that describe the interdependence of the four fields above. In a linear, isotropic, homogeneous, and reciprocal medium, these can be expressed by three scalar dimensionless quantities, the relative permittivity $\epsilon(k_0)$, the relative permeability $\mu(k_0)$, and the chirality parameter $\kappa(k_0)$~\cite{kristensson2016} as
\begin{align}\label{eq:constrel}
    \begin{pmatrix}
        \frac{1}{\epsilon_0} \bm D(\bm r, k_0) \\
        c \bm B(\bm r, k_0)
    \end{pmatrix}
    &=
    \begin{pmatrix}
        \epsilon(k_0) & \mathrm i \kappa(k_0) \\
        -\mathrm i \kappa(k_0) & \mu(k_0)
    \end{pmatrix}
    \begin{pmatrix}
        \bm E(\bm r, k_0) \\
        Z_0 \bm H(\bm r, k_0)
    \end{pmatrix}
    \,.
\end{align}
Thus, we arrive at
\begin{align}
    \nabla \times
    \begin{pmatrix}
        \bm E(\bm r, k_0) \\
        Z_0 \bm H(\bm r, k_0)
    \end{pmatrix}
    &=
    k_0
    \begin{pmatrix}
        \kappa(k_0) & \mathrm{i} \mu(k_0) \\
        -\mathrm i \epsilon(k_0) & \kappa(k_0)
    \end{pmatrix}
    \begin{pmatrix}
        \bm E(\bm r, k_0) \\
        Z_0 \bm H(\bm r, k_0)
    \end{pmatrix}
\end{align}
by combining \cref{eq:maxwell:curl,eq:constrel}. In the case of achiral materials, it becomes possible to decouple the electric and magnetic field equations by applying the curl twice, arriving at the Helmholtz wave equation. For non-zero $\kappa(k_0)$, the equations for these two fields can be decoupled by introducing the Riemann-Silberstein fields $\sqrt{2}\bm G_\pm(\bm r, k_0) = \bm E(\bm r, k_0) \pm \ii Z_0 Z(k_0) \bm H(\bm r, k_0)$~\cite{silberstein1907,lakhtakia1994} that then need to fulfill the differential equation
\begin{align}\label{eq:maxwell:g}
    \nabla \times \bm G_\pm(\bm r, k_0)
    = \pm k_\pm(k_0) \bm G_\pm(\bm r, k_0)\,,
\end{align}
where $Z(k_0) = \sqrt{\tfrac{\mu(k_0)}{\epsilon(k_0)}}$ is the relative wave impedance and $k_\pm(k_0) = k_0 (\sqrt{\epsilon(k_0)\mu(k_0)} \pm \kappa(k_0))$ is the wave number in the medium for each helicity. Applying the curl twice, we also obtain the Helmholtz equation for $\bm G_\pm(\bm r, k_0)$
\begin{equation}\label{eq:helmholtz}
    \left(\nabla^2 + k_\pm^2(k_0)\right)
    \bm G_\pm(\bm r, k_0)
    = 0
    \,.
\end{equation}
However, \cref{eq:maxwell:g} is actually stricter than the Helmholtz equation, i.e., not all solutions of the Helmholtz equation necessarily solve \cref{eq:maxwell:g}. However, starting from the solution theory developed for the Helmholtz equation, suitable solutions can be constructed easily, as shown in the following.

A set of transverse solutions for the vector Helmholtz equation obtained by applying the curl twice can be derived from solutions of the scalar Helmholtz equation $f_\nu(\bm r, k)$~\cite{morse1953}. These solutions are assumed to have a general index $\nu$. Then, the functions
\begin{subequations}\label{eq:helmholtz:sol:general}
    \begin{eqnarray}
        \bm M_\nu(\bm r, k)
        &
        = \nabla \times \left(\bm w f_\nu(\bm r, k)\right) \\
        \bm N_\nu(\bm r, k)
        &
        = \frac{\nabla}{k} \times \bm M_\nu(\bm r, k)
    \end{eqnarray}
\end{subequations}
are transverse solutions, where $\bm w$ is a pilot vector depending on the chosen basis set. For the planar and cylindrical case, it is $\bm w = \ez$, and in the spherical case, we use $\bm w = \bm r$. The solutions $\bm M_\nu(\bm r, k)$ are tangential to the surface defined by $\bm w$. When used for the electric field, which is possible for achiral media, they are called transverse electric (TE) waves. The solutions $\bm N_\nu(\bm r, k)$ are called transverse magnetic (TM). Note that the transversality always refers to the direction of $\bm w$. We will refer to the TE/TM modes also as parity modes. Solutions of well-defined helicity can be constructed by
\begin{align}\label{eq:anu}
    \bm A_{\nu,\pm}(\bm r, k) = \frac{\bm N_\nu(\bm r, k) \pm \bm M_\nu(\bm r, k)}{\sqrt{2}}
\end{align}
from the parity solutions. They then allow to expand the field
\begin{align}
    \bm G_\pm(\bm r, k_\pm(k_0)) = \sum_\nu a_{\nu,\pm} \bm A_{\nu,\pm}(\bm r, k_\pm(k_0))
\end{align}
in modes of positive or negative helicity. Inverting the definitions of these fields $\bm G_\pm(\bm r, k_\pm(k_0))$ leads to an expansion of the electric and magnetic fields in modes of well-defined helicity.

The first set of concrete solutions are PWs. The scalar solution $f_{\bm{\hat k}}(\bm r, k) = \ee^{\ii \bm{k r}}$ is indexed by the three Cartesian components of the normalized wave vector $\bm k = k \bm{\hat k}$. The application of \cref{eq:helmholtz:sol:general} results in
\begin{subequations}\label{eq:vpw}
    \begin{align}
        \bm M_{\bm{\hat k}}(\bm r, k)
        &
        =
        \ii\frac{k_y \ex - k_x \ey}{\sqrt{k_x^2 + k_y^2}}
        \ee^{\ii \bm{k r}}
        =
        -\ii \ephi_{\bm{\hat k}}
        \ee^{\ii \bm{k r}}
        \\
        \bm N_{\bm{\hat k}}(\bm r, k)
        &
        =
        \frac{
            -k_x k_z \ex - k_y k_z \ey + (k_x^2 + k_y^2) \ez
        }{
            k \sqrt{k_x^2 + k_y^2}
        }
        \ee^{\ii \bm{k r}}
        =
        -\etheta_{\bm{\hat k}}
        \ee^{\ii \bm{k r}}\,,
    \end{align}
\end{subequations}
where we divided the functions by $\sqrt{\smash[b]{k_x^2 + k_y^2}}$ to normalize the resulting modes. As shown on the right-hand side, these modes can be written compactly by using unit vectors of spherical coordinates (see \ref{app:coords}).

The CWs can be obtained from the scalar wave solutions $f_{k_z m}^{(n)}(\bm r, k) = Z_m^{(n)}(k_\rho \rho) \ee^{\ii m \phi + \ii k_z z}$, that are indexed by the wave's z-component $k_z$ and the azimuthal index $m \in \mathbb{Z}$. The vector $\bm r$ is expressed in cylindrical coordinates. Similarly, we define $k_\rho = \sqrt{\smash[b]{k^2 - k_z^2}}$. The function $Z_m^{(n)}$ can be the Bessel or Hankel functions of the first or second kind. As usual for T-matrices, we choose as the two independent solutions the Bessel functions of the first kind $Z_m^{(1)} = J_m$, that are regular in the whole space, and the Hankel functions of the first kind $Z_m^{(3)} = H_m^{(1)}$, that fulfill the radiation condition. After applying \cref{eq:helmholtz:sol:general} the basis set for CWs is
\begin{subequations}
    \begin{align}
        \bm M_{k_z m}^{(n)}(\bm r, k)
        &=
        \left(
            \ii m
            \frac{Z_m^{(n)}(k_\rho \rho)}{k_\rho \rho}
            \erho
            -
            {Z_m^{(n)}}'(k_\rho \rho)
            \ephi
        \right)
        \ee^{\ii m \phi + \ii k_z z}
        \\
        \bm N_{k_z m}^{(n)}(\bm r, k)
        &=
        \Bigg(
            \ii \frac{k_z}{k}
            {Z_m^{(n)}}'(k_\rho \rho)
            \erho
            -
            \frac{k_z m}{k}
            \frac{Z_m^{(n)}(k_\rho \rho)}{k_\rho \rho}
            \ephi
        +
        \frac{k_\rho}{k}        Z_m^{(n)}(k_\rho \rho)
        \ez
        \Bigg)
        \ee^{\ii m \phi + \ii k_z z}
        \,.
    \end{align}
\end{subequations}
As in the case of PWs, we normalize the fields. Here, this is done by dividing through $k_\rho$.

Finally, in analogy to the derivation of the CWs, the SWs can be derived. The scalar solution is $f_{lm}(\bm r, k) = z_l^{(n)}(k r) Y_{lm}(\theta, \phi)$, where the functions $z_l^{(n)}$ are the spherical Bessel and Hankel functions, analogously defined to the Bessel and Hankel functions for the CWs. The spherical harmonics $Y_{lm}(\theta, \phi)$ (\ref{app:sphharm}) have the indices $l \in \mathbb{N}$, where $l = 0$ corresponding to monopoles is not included for transverse waves, and $m \in \{-l, -l + 1, \dots, l\}$. With \cref{eq:helmholtz:sol:general} using $\bm w = \bm r$ the SWs are
\begin{subequations}
    \begin{align}
        \bm M_{lm}^{(n)} (\bm r, k)
        &=
        N_{lm}
        \left(
            \ii \pi_{lm}(\theta)
            \etheta
            -
            \tau_{lm}(\theta)
            \ephi
        \right)
        \ee^{\ii m \phi}
        z_l(k r)
        = \bm X_{lm}(\theta, \phi) z_l(kr)
        \\
        \bm N_{lm}^{(n)} (\bm r, k)
        &=
        \Bigg[
            \ii
            \sqrt{l (l + 1)}
            Y_{lm}(\theta, \phi)
            \frac{z_l(kr)}{kr}
            \er
            +
            \left({z_l^{(n)}}'(kr) + \frac{z_l^{(n)}(kr)}{kr}\right)
            \er \times \bm X_{lm}(\theta, \phi)
        \Bigg]\,,
    \end{align}
\end{subequations}
where we use the prefactor
\begin{equation}
    N_{lm}
    =
    \ii
    \sqrt{
        \frac{(2l + 1)}{4\pi l (l + 1)}
        \frac{(l - m)!}{(l + m)!}
    }
\end{equation}
and the angular functions
\begin{subequations}
    \begin{align}
        \pi_{lm}(\theta)
        &=
        \frac{m P_l^m(\cos \theta)}{\sin \theta}
        \\
        \tau_{lm}(\theta)
        &=
        \frac{\partial P_l^m(\cos \theta)}{\partial \theta}
    \end{align}
\end{subequations}
are derived from the associated Legendre polynomials (\ref{app:sphharm}).
With the PWs, the CWs, and the SWs defined, we can now use the properties of these functions to formulate the multi-scattering equations using the T-matrix method. For the PWs, we use an S-matrix description.

\subsection{T-matrix method in the presence and absence of periodic boundaries}
\label{sec:scattering:tmatrix}

The T-matrix method has been proven to be a suitable approach to efficiently perform electromagnetic scattering calculations for various systems. At its core, the T-matrix method relies on a set of functions that form a complete basis to Maxwell's equations together with the chosen constitutive relations, with -- among others -- two properties: First, the functions can be separated into incident and scattered waves and, second, simple translation properties of the individual waves, such that multiple scattering calculations can be done efficiently, are known. Typically, these properties are associated with the SWs. However, it is equally possible to apply this concept to CWs. Due to the analytically known properties of these functions, not only the scattering coefficients but also various quantities, such as scattering and extinction cross-sections, circular dichroism, or duality breaking, can be computed highly efficiently.

For the T-matrix method, the total electric field is separated into incident and scattered modes that can be expanded as
\begin{align}
    \begin{split}
        \bm E_\text{inc}(\bm r, k_0)
        =
        \sum_\nu
        \sum_{\lambda = \pm 1}
        a_{\nu\lambda} \bm A_{\nu,\lambda}^{(1)}(\bm r, k_\lambda(k_0))
        \\
        \bm E_\text{sca}(\bm r, k_0)
        =
        \sum_\nu
        \sum_{\lambda = \pm 1}
        p_{\nu \lambda} \bm A_{\nu,\lambda}^{(3)}(\bm r, k_\lambda(k_0))
    \end{split}
\end{align}
using the previously derived solutions in the helicity basis. In the case of vanishing chirality parameters, the fields can be expanded equally well in the parity basis of $\bm M_{\nu}^{(n)}(\bm r, k(k_0))$ and $\bm N_{\nu}^{(n)}(\bm r, k(k_0))$. \treams{} has both solution sets implemented, but for brevity, we will restrict the introduction here to the helicity modes. The sum over $\nu$ is a shorthand for the sum over $l$ and $m$ in the case of SWs and for $m \in \mathbb{Z}$ and $k_z \in \mathbb{R}$ for CWs. The continuous index $k_z$ needs, in principle, an integral expression in the expansion. However, we will restrict the application of CWs to systems with periodicity along $z$ where only discrete diffraction orders couple. In both cases, the sum has to be truncated to obtain a finite matrix. By choosing a large enough value, the error introduced by this truncation can be controlled.
For SWs, we, therefore, assume a finite scatterer size, such that a finite number of multipoles is sufficient and only waves with $l \leq l_\text{max}$ are used. For CWs, we first assume periodicity in the z-direction. Then, the sum over $k_z$ only includes values that differ by multiples of the reciprocal lattice. Thus, for objects with a finite extent in the x- and y-directions, the sum over diffraction orders $k_z$ and the sum over $|m| \leq m_\text{max}$ can be truncated. The finite number of incident and scattered field coefficients constitute then the vectors $\bm a$ and $\bm p$. Now, for a linear response, a relationship between the incident and scattered field can be expressed by the T-matrix $\mathbf{T}$ through
\begin{align}
    \bm p = \mathbf{T} \bm a\,.
\end{align}
So, the T-matrix can be seen as a full description of the electromagnetic properties at a specified frequency. It can be obtained analytically for spheres in SWs and infinitely long cylinders in CWs, which can consist of multiple concentric shells~\cite{bohren1998a,moroz2005a,shang2016a}. The calculation of these coefficients is included in \treams. For more complicated objects, the T-matrix can be computed by a range of numerical methods~\cite{doicu1997,mackowski2002a,fruhnert2017c,demesy2018}.

To describe the interaction between $N$ particles with T-matrices $\mathbf{T}_i$ at positions $\bm r_i$ with $i \in \{1, \dots, N\}$, we have to add to the primary incident field $\bm a_i$ also the scattered fields of all other particles by
\begin{align}\label{eq:tmat:cluster}
    \bm p_i = \mathbf{T}_i \left[ \bm a_i + \sum_{j \neq i} \mathbf{C}^{(3)}(\bm r_i - \bm r_j) \bm p_j\right]\,,
\end{align}
where $\mathbf{C}^{(3)}(\bm r_i - \bm r_j) = \mathbf{C}^{(3)}_{i,j}$ contains the translation coefficients (\ref{app:transl}) along $\bm r_i - \bm r_j$~\cite{cruzan1962,stein1961,tsang1985}. This set of linear equations can be solved in the local basis, where the field expansions at all positions $\bm a_i$ and $\bm p_i$ are combined in one vector $\bm a_\text{local}$ and $\bm p_\text{local}$~\cite{suryadharma2017a}. The total scattered field is then the superposition of all individual scattered fields.
Using the translation properties of the SWs, we can write the multi-scattering equation as
\begin{align}\label{eq:tmat:local:alt}
    \bm p_\text{local}
    =
    \mathbf{T}_\text{diag}
    \left[
        \bm a_\text{local}
        +
        \mathbf{C}^{(3)}
        \bm p_\text{local}
    \right]\,,
\end{align}
where $\mathbf{T}_\text{diag}$ is the block-diagonal matrix containing the T-matrices of all particles 
\begin{align}\label{eq:cmat}
    \mathbf{T}
    =
    \begin{pmatrix}
        \mathbf{T}_1 & 0 & \dots & 0 \\
        0 & \mathbf{T}_2 & \ddots & \vdots \\
        \vdots & \ddots & \ddots & 0 \\
        0 & \dots & 0 & \mathbf{T}_N
    \end{pmatrix}
    \quad\text{and}\quad
    \mathbf{C}^{(3)}
    =
    \begin{pmatrix}
        0 & \mathbf{C}^{(3)}_{1,2} & \dots & \mathbf{C}^{(3)}_{1,N} \\
        \mathbf{C}^{(3)}_{2,1} & 0 & \ddots & \vdots \\
        \vdots & \ddots & \ddots & \mathbf{C}^{(3)}_{N-1,N} \\
        \mathbf{C}^{(3)}_{N,1} & \dots & \mathbf{C}^{(3)}_{N,N-1} & 0
    \end{pmatrix}
\end{align}
expands for each particle the scattered fields from all particles in regular waves using the translation coefficients.
Equation~(\ref{eq:tmat:local:alt}) can be rewritten as 
\begin{align}\label{eq:tmat:local}
    \bm p_\text{local}
    =
    \underbrace{
        \left[\mathds{1} - \mathbf{T}_\text{diag}\mathbf{C}^{(3)}\right]^{-1} \mathbf{T}_\text{diag}
    }_{=\mathbf{T}_\text{local}}
    \bm a_\text{local}
\end{align}
to obtain the local T-matrix for the interacting particles. That local T-matrix for the interacting particles can already directly be used to compute quantities such as the scattering cross-section without needing to convert it to the global basis. The expansion of the incident and scattered field is taken at $N$ positions.
By expressing the incident fields at each particle in terms of the expansion at the origin $\bm a$ with $\bm r_0 = 0$
\begin{align}
    \bm a_\text{local} = \begin{pmatrix}
        C^{(1)}_{1, 0} & \dots & C^{(1)}_{N, 0}
    \end{pmatrix}
    \bm a
\end{align}
and, likewise, summing the scattered fields from all particles to express them in an expansion at the origin
\begin{align}
    \bm p = \begin{pmatrix}
        C^{(1)}_{0, 1} & \dots & C^{(1)}_{0, N}
    \end{pmatrix}
    \bm p_\text{local}\,,
\end{align}
we can obtain a global T-matrix
\begin{align}
    \mathbf{T}
    =
    \begin{pmatrix}
        C^{(1)}_{0, 1} & \dots & C^{(1)}_{0, N}
    \end{pmatrix}
    \mathbf{T}_\text{local}
    \begin{pmatrix}
        C^{(1)}_{1, 0} & \dots & C^{(1)}_{N, 0}
    \end{pmatrix}
\end{align}
from the local T-matrix. In practice, \treams{} does not distinguish between local and global T-matrices: the global T-matrix is treated as a local T-matrix in the special case that only one origin is present.

Next, we want to consider periodic boundaries. We assume that there are $N$ particles per unit cell, with $\bm r_i - \bm r_j$ pointing from particle $j$ to particle $i$ in the same unit cell. The $d$-dimensional lattice is defined by the basis vectors $\bm u_i$ with $i \in \{1, \dots, d\}$ and consists of the set $\Lambda_d = \big\{\sum_{i=1}^d n_i \bm u_i | n_i \in \mathbb{Z}\big\}$. We assume that the illumination of the array has a fixed phase relation along the direction of the lattice given by the vector $\bm k_\parallel$. Thus, the phase difference between unit cells of distance $\bm R \in \Lambda_d$ is $\ee^{\ii \bm k_\parallel \bm R}$.
We select one unit cell as our reference that is placed at the origin by convention. Then, the scattered fields amplitudes $\bm{\tilde p}_i$ are defined analogously to \cref{eq:tmat:cluster} but now including a sum over all lattice sites, by
\begin{align}
    \bm{\tilde p}_i
    = \mathbf{T}_i
    \left[
        \bm{\tilde a}_i
        +
        \sum_{j = 1}^N
        \sideset{}{'}{\sum}_{\bm R \in \Lambda_d}
        \mathbf{C}^{(3)}(\bm r_i - \bm r_j - \bm R) \bm{\tilde p}_{j,\bm R}
    \right]
    \,.
\end{align}
The prime next to the lattice sum indicates the omission of the self-interaction term with $\bm R = 0$ if $i = j$.
Making use of the fixed phase relation $\bm{\tilde p}_{j, \bm R} = \ee^{\ii \bm k_\parallel \bm R} \bm{\tilde p}_j$, we can convert this expression to
\begin{align}\label{eq:tmat:lattice}
    \bm{\tilde p}_i
    = \mathbf{T}_i
    \Bigg[
        \bm{\tilde a}_i
        +
        \sum_{j = 1}^N
        \underbrace{\sideset{}{'}{\sum}_{\bm R \in \Lambda_d} \mathbf{C}^{(3)}(\bm r_i - \bm r_j - \bm R) \ee^{\ii \bm k_\parallel \bm R}}_{=\mathbf{\tilde{C}}^{(3)}_{ij}}
        \bm{\tilde p}_j
    \Bigg]\,,
\end{align}
with only quantities from the reference unit cell.

\begin{figure}
    \centering
    \includegraphics[width=.9\linewidth]{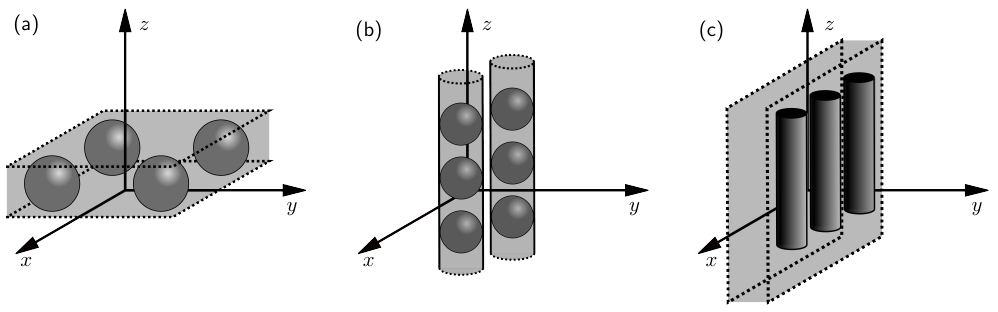}
    \caption{The domains of validity for different expansion and lattices. For objects described by a T-matrix, i.e., the incident and scattered fields are expanded in SWs, the expansion is only valid outside the circumscribing sphere~\cite{peterson1973a}. Then, for a 2D lattice in the x-y-plane of T-matrices, the expansion in PWs is valid above and below the planes that enclose the lattice and the scatterers, as shown in panel (a). Panel (b) shows the domain of validity when a 1D lattice of SW expansions along the z-axis is expanded in a CW. As shown, a local T-matrix in SWs can be expanded in a local CW basis, where each position around which the field is expanded has its own domain of validity. The CWs are only valid in the domain outside of a circumscribing cylinder. Similarly to the expansion of SWs in PWs in panel (a), the expansion of CWs on a 1D lattice along the x-axis in PWs is only valid outside the planes that enclose the lattice.}
    \label{fig:validity}
\end{figure}

The lattice sums over translation coefficients converge notoriously slow~\cite{linton2010}. However, they can be considerably accelerated using the Ewald summation technique~\cite{ewald1921a,linton2010,moroz2006a,kambe1968,eyert2012}. For each of the basis sets, SWs and CWs, combined with each possible lattice dimension, they have to be evaluated individually. An additional difficulty that arises in these sums is the inclusion of complex unit cells, which additionally require shifts within the unit cell. We derived these lattice sums in a unified manner previously~\cite{beutel2023}. All the lattice sums are implemented in \treams. These lattice sums assume a default orientation of the lattices. 2D lattices of SWs are placed in the x-y-plane, 1D lattices of SWs are placed along the z-direction, and 1D lattices of CWs are placed along the x-direction as shown in \cref{fig:validity}.
Using the definition of the translation matrix with periodic boundaries $\mathbf{\tilde C}^{(3)}$ analogous to \cref{eq:cmat} but now with the diagonal blocks generally being non-zero and containing the interaction of the particle with its periodic equivalents, we can write 
\begin{align}
    \bm{\tilde{p}}_\text{local}
    =
    \left[
        \mathds{1}
        -
        \mathbf{T}_\text{diag}
        \mathbf{\tilde{C}}^{(3)}
    \right]^{-1}
    \mathbf{T}_\text{diag}
    \bm{\tilde{a}}_\text{local}\,,
\end{align}
which is formally quite similar to \cref{eq:tmat:local}. However, the matrix $\mathbf{\tilde C}^{(3)}$ includes a sum over lattice sites, that the incident field is assumed to fulfill the phase relation defined by $\bm k_\parallel$, and that the total scattered field is not only obtained after summing over all particle positions but also all lattice sites. To find lattice modes supported in the lattice without external illumination, we can solve
\begin{align}
    \left[
        \mathds{1}
        -
        \mathbf{T}_\text{diag}
        \mathbf{\tilde{C}}^{(3)}
    \right]
    \bm{\tilde{p}}_\text{local}
    = 0\,,
\end{align}
which in practice can be done by calculating eigenmodes with vanishing or sufficiently small eigenvalues.

\subsection{S-matrices for stratified structures\label{sec:scattering:smatrix}}

For PWs, we use an S-matrix description. This means, instead of incident and scattered modes, we use a separation into incoming and outgoing modes. These modes are defined concerning a reference plane that we assume to be perpendicular to $\ez$. Similarly to the CWs, we will assume periodicity in the x-y-plane. Then, we can make a slight modification of the mode definition. In \cref{eq:vpw}, the PWs are indexed by the three components of $\bm{\hat k}$, now we use the following definition that is more suitable for systems with periodicity in the x-y-plane later: We index the modes with $\bm k_\parallel = k_x \ex + k_y \ey$ and additionally a direction $d = \pm 1$ that defines the sign of the remaining component $k_z = \pm\sqrt{\smash[b]{k^2 - k_\parallel^2}}$. We chose the same symbol $\bm k_\parallel$ as in \cref{eq:tmat:lattice} because the parallel component here can be immediately used in that equation as the phase difference between unit cells. The incoming fields are
\begin{align}
    \bm E_\text{in}(\bm r, k_0) = \sum_{\bm k_\parallel} \sum_{\lambda = \pm 1} a_{\bm k_\parallel d \lambda} \bm A_{\bm k_\parallel d, \lambda}(\bm r, k_\lambda(k_0))\,,
\end{align}
with $d = \pm 1$ for $z \lessgtr 0$ and the outgoing fields are
\begin{align}
    \bm E_\text{out}(\bm r, k_0) = \sum_{\bm k_\parallel} \sum_{\lambda = \pm 1} b_{\bm k_\parallel d \lambda} \bm A_{\bm k_\parallel d, \lambda}(\bm r, k_\lambda(k_0))\,.
\end{align}
Therefore, to describe the scattering of a particular PW in a 2D periodic system, it is sufficient to reduce the values of $\bm k_\parallel$ from a continuous spectrum to the countable set of diffraction orders indicated by the sum symbol. Furthermore, we can truncate the sum to a finite number of modes, neglecting only highly evanescent contributions to the fields. After sorting the expansion coefficients into four sets of modes $\bm a_\uparrow$, $\bm a_\downarrow$, $\bm b_\uparrow$, and $\bm b_\downarrow$, we can describe the scattering with
\begin{align}\label{eq:smatrix}
    \begin{pmatrix}
        \bm b_\uparrow \\
        \bm b_\downarrow \\
    \end{pmatrix}
    =
    \underbrace{
    \begin{pmatrix}
        \mathbf{S}_{\uparrow\uparrow} & \mathbf{S}_{\uparrow\downarrow} \\
        \mathbf{S}_{\downarrow\uparrow} & \mathbf{S}_{\downarrow\downarrow} \\
    \end{pmatrix}
    }_\mathbf{S}
    \begin{pmatrix}
        \bm a_\uparrow \\
        \bm a_\downarrow \\
    \end{pmatrix}
    \,.
\end{align}
To avoid confusion between the direction $d = \pm 1$ and the helicity $\lambda$, we used $\uparrow$ and $\downarrow$ to indicate the direction. We implicitly assume that the scattering takes place at the $z = 0$ plane. Except for planar interfaces, this is obviously only a model description. The domain of validity of the field expansion is then only above and below the planes that encase the scatterers geometrically, as shown in \cref{fig:validity}(a). The two diagonal blocks contain the coefficients for transmission. The off-diagonals contain the coefficients for reflection.

Two systems described by the matrices $\mathbf{S}^I$ and $\mathbf{S}^{II}$ can be stacked, where the resulting S-matrix is given by
\begin{subequations}
\begin{align}
    \mathbf{S}_{\uparrow\uparrow}
    &=
    \mathbf{S}_{\uparrow\uparrow}^{II}
    \left[1 - \mathbf{S}_{\uparrow\downarrow}^{I}\mathbf{S}_{\downarrow\uparrow}^{II}\right]^{-1}
    \mathbf{S}_{\uparrow\uparrow}^{I}
    \\
    \mathbf{S}_{\uparrow\downarrow}
    &=
    \mathbf{S}_{\uparrow\uparrow}^{II}
    \left[1 - \mathbf{S}_{\uparrow\downarrow}^{I}\mathbf{S}_{\downarrow\uparrow}^{II}\right]^{-1}
    \mathbf{S}_{\uparrow\downarrow}^{I}
    \mathbf{S}_{\downarrow\downarrow}^{II}
    +
    \mathbf{S}_{\uparrow\downarrow}^{II}
    \\
    \mathbf{S}_{\downarrow\downarrow}
    &=
    \mathbf{S}_{\downarrow\downarrow}^{I}
    \left[1 - \mathbf{S}_{\downarrow\uparrow}^{II}\mathbf{S}_{\uparrow\downarrow}^{I}\right]^{-1}
    \mathbf{S}_{\downarrow\downarrow}^{II}
    \\
    \mathbf{S}_{\downarrow\uparrow}
    &=
    \mathbf{S}_{\downarrow\downarrow}^{I}
    \left[1 - \mathbf{S}_{\downarrow\uparrow}^{II}\mathbf{S}_{\uparrow\downarrow}^{I}\right]^{-1}
    \mathbf{S}_{\downarrow\uparrow}^{II}
    \mathbf{S}_{\uparrow\uparrow}^{I}
    +
    \mathbf{S}_{\downarrow\uparrow}^{I}\,,
\end{align}
\end{subequations}
taking the multi-reflection between the two objects described by the S-matrices into account. For repeating structures along the z-direction, the layer doubling technique is highly efficient in calculating the response~\cite{pendry1974}.
Furthermore, it is conveniently possible to calculate the band structure of periodic systems along the z-direction from the S-matrix of a single layer \cite{stefanou1998,pendry1974}.

\subsection{Further Operators\label{sec:scattering:ops}}

Up to now, we already considered few operators, in the sense that they perform a particular action when applied to, e.g., a field expansion. The T-matrix or S-matrix can be seen as an operator, also the matrices that transform the scattered fields back into incident fields. Here, we will discuss several other common operations implemented in \treams. These are translations and rotations of objects, the expansion of a field in another basis, and the expansion of the scattered fields in a lattice in another basis.

We already introduced the expansion of scattered fields in incident fields for CWs and SWs $\mathbf{C}^{(3)}$ and the matrices to expand those fields in the same type $\mathbf{C}^{(1)}$, i.e., expand incident fields at some point into incident fields at another point and, analogously, for the expansion of scattered fields in scattered fields. A missing part, especially for connecting different basis sets, is the expansions in one basis set into other basis sets. First, we have the expansion of PWs in SWs~\cite{mishchenko1996}
\begin{align}\label{eq:ex:pwsw}
    \bm A_{\bm k, \pm}(\bm r, k)
    =
    4 \pi
    \sum_{l = 1}^\infty
    \sum_{m = -l}^l
    N_{lm}
    \ii^l
    \left(
        \tau_{lm}(\cos\theta_{\bm k})
        \pm
        \pi_{lm}(\cos\theta_{\bm k})
    \right)
    \ee^{-\ii m \phi_{\bm k}}
    \bm A_{lm,\pm}^{(1)}(\bm r, k)
\end{align}
and of PWs in CWs~\cite{frezza2018}
\begin{align}\label{eq:ex:pwcw}
    \bm A_{\bm k, \pm}(\bm r, k)
    =
    \sum_{m = -\infty}^\infty
    \ii^m
    \ee^{-\ii m \phi_{\bm k}}
    \bm A_{k_z m,\pm}^{(1)}(\bm r, k)
    \,.
\end{align}
These expansions can be obtained from the scalar plane wave expansions and then applying the respective operations from \cref{eq:helmholtz:sol:general}. The PWs are regular in the whole space, so their expansion consists only of regular CWs or regular SWs. A benefit of the helicity basis is apparent in these equations: opposite helicities are not mixed by these operations. For a truncated expansion, the coefficients can be used in matrices to describe the change of basis. By equating and comparing the coefficients from the two PWs expansions, we get the missing expansion~\cite{han2007}
\begin{align}\label{eq:ex:cwsw}
    \bm A_{k_z m,\pm}^{(1)}(\bm r, k)
    =
    \sum_{l = \max(|m|, 1)}^\infty
    4 \pi
    N_{lm}
    i^{l - m}
    \left(
        \tau_{lm}(\cos\theta_{\bm k})
        \pm
        \pi_{lm}(\cos\theta_{\bm k})
    \right)
    \bm A_{lm,\pm}^{(1)}(\bm r, k)
\end{align}
of CWs in SWs.
The inverse transformations of these expressions can be derived from the Fourier transforms of the scalar cylindrical and spherical waves~\cite{wittmann1988}. They result in continuous spectra and, thus, require discretization of the spectrum to describe them with simple matrices.

We find a convenient connection between the chosen basis sets for periodic boundary conditions that can simplify and speed-up computations. As mentioned earlier, each type of solution has a geometric symmetry where it is most beneficial. SWs are well-suited for objects of finite size, CWs are well-suited for objects that are periodic along one axis and of finite extent in the remaining two directions, and PWs can be efficiently used for 2D periodic structures. We can easily find transformations between basis sets by taking these natural symmetries of the different solution sets as guidance. For example, summing singular SWs placed on a 1D lattice along the z-axis leads to CWs. In such a case, the typical trade-off is between the speed of the calculation and the convergence area of the solution, as shown in \cref{fig:validity}. While using SWs together with lattice sums provides convergence in the whole area outside the circumscribing spheres of each particle, the calculation, although considerably accelerated by Ewald's method, is comparably slow compared to the use of CWs. This basis set is inherently efficient for 1D periodic structures. However, the solution now only converges outside the circumscribing cylinder. In effect, these two methods complement each other. Analogously to the described transition from 1D lattices of SWs to CWs, it is possible to find transitions from SWs on a 2D lattice to PWs and from CWs on a 1D lattice to PWs. In the latter case, the inherent 1D periodicity of the CWs in one direction, in addition to their arrangement on a 1D lattice in a perpendicular direction, leads to the required 2D periodicity for the PWs. Moreover, under certain conditions, it is possible that by changing the basis, it becomes possible to expand the range of validity in certain parts of the domain~\cite{egel2017a}.

By using the Fourier space representation of the CWs and the PWs together with Poisson's equation for lattice sums (\ref{app:expandlattice}), we obtain
\begin{align}\label{eq:exl:swcw}
    \sum_{\bm R \in \Lambda_1}
    \bm A_{lm,\lambda}^{(3)}(\bm{r} - \bm{R}, k)
    \ee^{\ii \bm k_\parallel \bm R}
    &=
    -\frac{\ii N_{lm} \pi}{a k \ii^{l - m}}
    \sum_{\bm G \in \Lambda_1^\ast}
    \left(
        \tau_{lm}(\theta_{\bm k}) 
        \pm \pi_{lm}(\theta_{\bm k}) 
    \right)
    \bm A_{G+k_\parallel m, \lambda}^{(3)}(\bm r, k)\,,
\end{align}
where, by assuming the lattice $\Lambda_1 = \big\{n a \ez | n \in \mathbb{Z}\big\}$ to be in the z-direction with pitch $a$, we have on the right-hand side of the equation the sum over diffraction orders $\bm G = G\ez \in \big\{n \tfrac{2\pi}{a}\ez | n \in \mathbb{Z}\big\}$. Furthermore, we have $\bm k_\parallel = k_\parallel \ez$. This geometry implies $\cos\theta_{\bm k} = \tfrac{k_\parallel + G}{k}$.

Similarly, we can transform a sum on a 2D lattice in the x-y-plane to PWs by transforming them with the expansion
\begin{align}\label{eq:exl:swpw}
    \sum_{\bm R \in \Lambda_2}
    \bm A_{lm, \lambda}^{(3)}(\bm{r} - \bm{R}, k)
    \ee^{\ii \bm k_\parallel \bm R}
    &=
    -\frac{\ii N_{lm} 2\pi}{A k^2 \ii^l}
    \sum_{\bm G \in \Lambda_2^\ast}
    \frac{\ee^{\ii m \phi_{\bm k}}}{\sqrt{1 - \tfrac{(\bm k_\parallel + \bm G)^2}{k^2}}}
        \left(\tau_{lm}(\theta_{\bm k}) \pm \pi_{lm}(\theta_{\bm k}) \right)
    \bm A_{\bm G + \bm k_\parallel, d, \lambda}(\bm{r}, k)\,,
\end{align}
where $A$ is the area of one unit cell and $\cos\theta_{\bm k} = d\tfrac{\sqrt{\smash[b]{k^2 - (\bm k_\parallel + \bm G)^2}}}{k}$. Lastly, there is the transformation of CWs on a 1D lattice along the x-direction into PWs
\begin{align}\label{eq:exl:cwpw}
    \sum_{\bm R \in \Lambda_1}
    \bm A_{k_z m, \lambda}(\bm r - \bm R, k) \ee^{\ii \bm k_\parallel \bm R}
    =
    \frac{2 (-\ii)^m}{a k}
    \sum_{\bm G \in \Lambda_1^\ast}
    \frac{\ee^{\ii m \phi_{\bm k}}}{\sqrt{1 - \tfrac{(\bm k_\parallel + \bm G)^2}{k^2}}}
    \bm A_{\bm{\hat k}, \lambda}(\bm r, k)\,,
\end{align}
with $\bm k = (k_\parallel + G)\ex + d \sqrt{\smash[b]{k^2 - (k_\parallel + G)^2 - k_z^2}} \ey + k_z \ez$
Note that in the last case, the periodicity is in the x-z-plane. To permute the labeling of the axes to the more conventional periodicity in the x-y-plane, the formula (\ref{app:perm})
\begin{align}
    \bm A_{\bm{{\hat{k}}'}, \lambda}(\bm r', k) = \frac{-k_x k_z + \lambda  \ii k k_y}{\sqrt{k_x^2 + k_y^2} \sqrt{k_y^2 + k_z^2}} \bm A_{\bm{\hat k}, \lambda}(\bm r, k)
\end{align}
can be used, where we have the transformation from $\bm r' = z \ex + x \ey + y \ez$ and $\bm k' = k_z \ex + k_x \ey + k_y \ez$ to $\bm r = x \ex + y \ey + z \ez$ and $\bm k = k_x \ex + k_y \ey + k_z \ez$.

Other common transformations are the rotations by Euler angles or translations along some direction. For SWs, rotations are represented by the Wigner-D matrix elements~\cite{dachsel2006a}. For PWs and CWs, currently, only rotations about the z-axis are implemented. Translations along a vector $\bm r$ for PWs only include a phase factor $\ee^{\ii \bm k \rm r}$. For CWs and SWs, the translation in the local basis are just block-diagonal matrices containing the translation coefficients $\bm C^{(1)}(\bm r)$. The last transformation to be explicitly mentioned here is the change of the polarization between helicity and parity basis, which is given by \cref{eq:anu}.

\section{Program overview}\label{sec:treams}

\begin{figure}
    \centering
    \includegraphics[width=.5\linewidth]{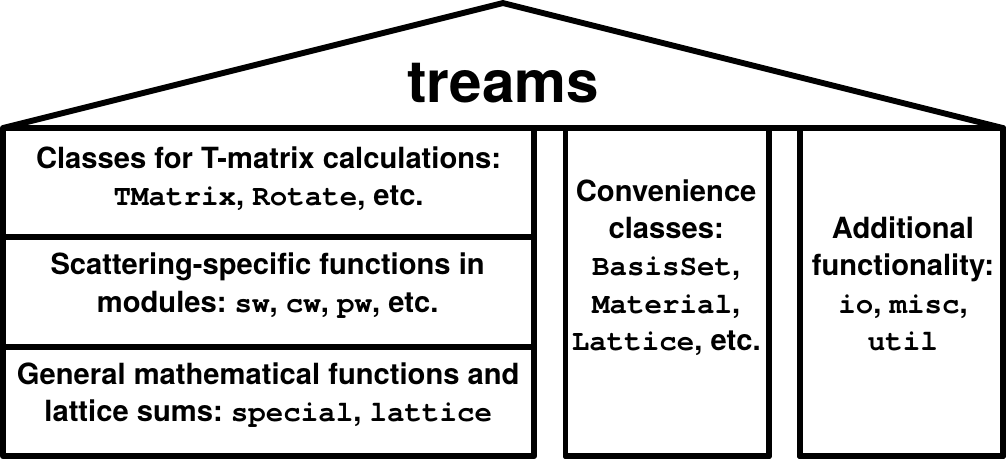}
    \caption{The general structure of \treams. The T-matrix calculations can be roughly separated into three levels. The focus for the lower levels was speed, and they are therefore implemented in Cython. They are generally available as \emph{numpy} universal functions, which integrates them neatly into the general framework for numerical linear algebra operations. The higher levels are more specifically aimed at scattering calculations and provide more tailored access to the underlying functions. On the highest level, the classes provide more convenience, functionality, and plausibility checks. This functionality is paired with several classes that support creating the necessary definitions of basis set, lattices, and other structures. The program is completed with functions, e.g., to load and store T-matrices.}
    \label{fig:structure}
\end{figure}

The program \treams{} is a Python package that provides functionality to perform electromagnetic scattering calculations using the T-matrix method. It aims to provide an accessible framework for fast calculations using the basis sets defined in the previous section and operations on them in combination with implementations of exponentially fast converging lattice sums. First, we will introduce the general structure of \treams, where on the one hand, we have fast and flexible functions for multi-scattering calculations, and on the other hand, a more high-level description of the T-matrices and possible operations on them that provide more convenience and plausibility checks of the arguments in functions. We describe the implemented operators and, eventually, the implemented classes for T-matrices and S-matrices.

\subsection{General structure\label{sec:treams:structure}}
A broad overview of \treams{} is shown in \cref{fig:structure}. One of the major parts of the program is, of course, directly related to electromagnetic scattering calculations. These form the first column supporting \treams. The second column is related to convenience classes that hold descriptions, e.g., of the lattices or materials. Lastly, there are modules with additional functionality to support, for example, loading or storing T-matrices into a file format based on the HDF5 format or the interface to \emph{numpy}~\cite{harris2020} arrays, which are one of the standard ways to store matrices and vectors in computer memory.

The first column, containing the electromagnetic scattering modules, classes, and functions, can be further separated into three levels. At the base, there are the implementations of mathematical functions that are not available from widely used packages such as \emph{scipy}~\cite{virtanen2020}: these include, for example, the different types of waves and their translation coefficients and associated functions such as the Wigner-3j symbols~\cite{xu1998}. In analogy to the submodule in \emph{scipy}, these functions are in a module called \texttt{special} (from special mathematical functions). The other part of this low-level implementation is the lattice sums and associated integrals. These basic methods are optimized for speed. Therefore, they are implemented in Cython, which gets converted to C-code and compiled, increasing the speed of the computations. These functions are then exposed in Python as \emph{numpy} universal functions. This makes those functions behave similarly to the functions of \emph{numpy} itself, such as the exponential function. For example, they are vectorized and broadcast when called with arrays as arguments. Furthermore, the implemented functions are also pure functions. Since those functions are available as independent modules, they can easily be reused in other programs. Moreover, a Cython-interface to those functions is provided via \texttt{special.cython\_special} and \texttt{lattice.cython\_lattice}, such that they can be reused in other compiled code.

The next higher level contains the modules \texttt{pw}, \texttt{cw}, \texttt{sw}, and \texttt{coeff}. They provide mathematical functions with a dedicated purpose, e.g., implementing the basis change coefficients with all prefactors. These functions are still provided as part of compiled universal functions with the same procedure as explained for the basic functions. This makes the separation of those two levels arbitrary to some point. However, by providing a coarse categorization by the basis set involved, they are intended to help find the relevant functions quickly. The module \texttt{coeff} offers functions to compute scattering coefficients, namely the Fresnel coefficients for planar interfaces and multilayers with chirality, the Mie coefficients for chiral spheres also generalized for spheres with additional spherical shells, and the scattering coefficients for infinitely long cylinders in the CW basis.

\begin{figure}
    \centering
    \includegraphics[width=.95\linewidth]{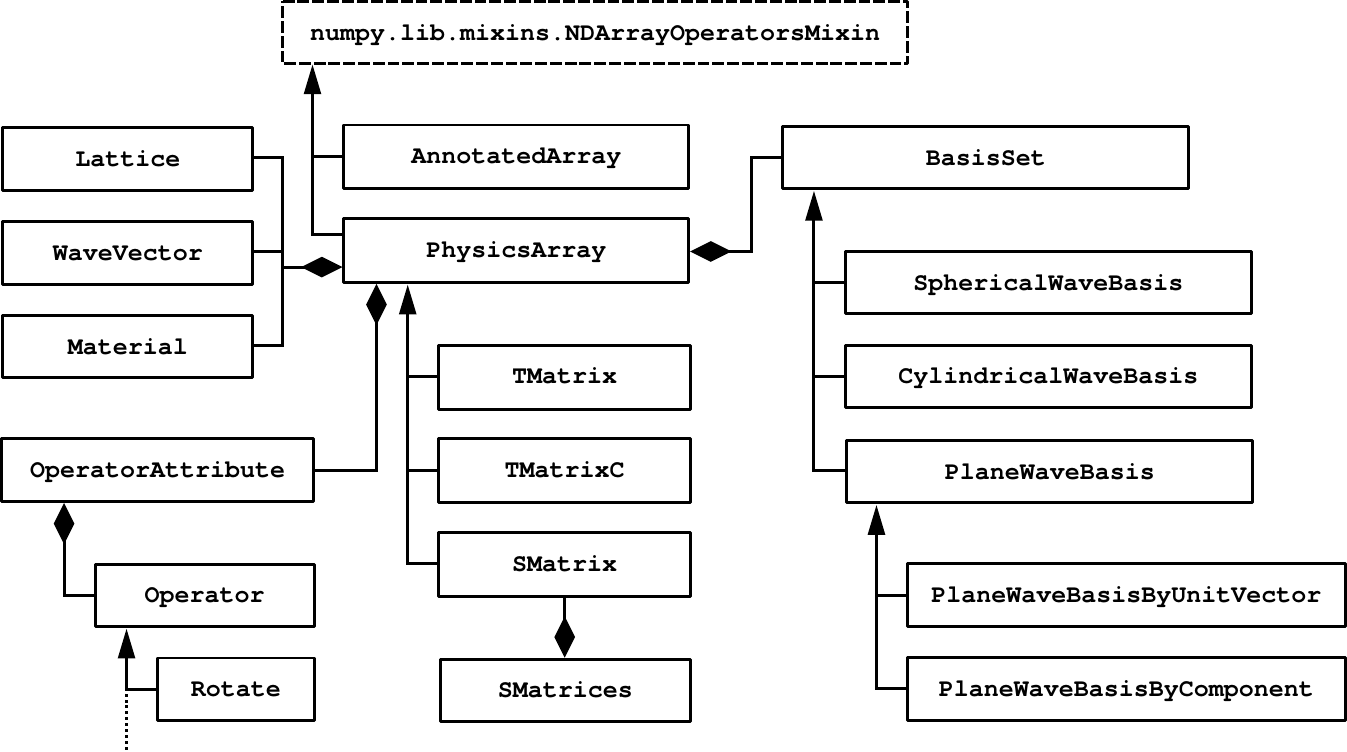}
    \caption{Major classes of \treams. The central column shows structures that behave mostly like \emph{numpy} arrays and adds additional functionality to it. The central class for most of the functionality is \texttt{PhysicsArray}. Instances of that class keep track of physically relevant information, such as the wave number, the basis set, the embedding material, etc. Some of these objects are implemented as classes themselves. Especially the class \texttt{BasisSet} is the parent to all available types of basis sets implemented. A range of operators is wrapped as attributes and composed to the class.}
    \label{fig:uml}
\end{figure}

The high-level classes provide an alternative interface to these functions through several classes. They interact closely with the functionality of classes listed in the other columns. A detailed unified modeling language (UML) style diagram of the relevant classes is shown in \cref{fig:uml}. Central to the calculations is the class \texttt{PhysicsArray}, which behaves like \emph{numpy} arrays by using interfaces provided by \emph{numpy} and deriving from its class \texttt{numpy.lib}\allowbreak\texttt{.mixins}\allowbreak\texttt{.NDArrayOperatorMixin}. The parent class \texttt{AnnotatedArray} implements a general way to store information, such as the underlying basis set, in each dimension of the array. In a \texttt{PhysicsArray}, these annotations are restricted to a defined set of physical parameters, which can then be exposed as attributes. Thus, a \texttt{PhysicsArray} holds, e.g., wave number, lattice, wave vector, material, or basis set information in a composition. Compositions are shown with the diamond symbol.

Another part composed of a \texttt{PhysicsArray} is called \texttt{OperatorAttribute}. They perform certain transformations on the values in the array and will be treated in \cref{sec:treams:ops}. Finally, on the right side of \cref{fig:uml}, we see the different types of basis sets: they obviously hold the SWs, the CWs, and the PWs. The latter are further separated into two possible descriptions. First, a PW basis can be described by a set of unit vectors $\bm{\hat k}$. Second, the wave can be described by two components and additionally the sign of the remaining third component, which is a convenient representation in the presence of a lattice and diffraction orders, as mentioned earlier.
The classes for T-matrices and S-matrices then derive from the \texttt{PhysicsArray}. They place even tighter restrictions on the annotations, e.g., the basis set has to be of a certain type. Most operations, like rotations, are then automatically implemented by the operators of the parent class \texttt{PhysicsArray}. Only specific functions, e.g., to solve the multi-scattering problem in \cref{eq:tmat:local} or to compute scattering and extinction cross-sections, are explicitly implemented for the T-matrices. We will discuss the operators and the classes \texttt{PhysicsArray}, \texttt{TMatrix}, and \texttt{SMatrix} now in more detail. 

\begin{table}
    \centering
    \caption{Available Operators}
    \begin{tabular}{ll}
        \toprule
         Operator name & Description \\
         \midrule
         \texttt{Rotate} & Rotate field by Euler angles \\
         \texttt{Translate} & Translate field by a vector \\
         \texttt{Expand} & Expand field in another basis/at other positions \\
         \texttt{ExpandLattice} & \makecell[l]{Expand field in another basis/at other positions\\assuming a periodic lattice} \\
         \texttt{Permute} & Permute the axes of PWs \\
         \texttt{ChangePoltype} & Switch between helicity and parity basis \\
         \texttt{EField}, \texttt{HField}, \dots & Evaluate the field at specified positions \\
         \bottomrule
    \end{tabular}
    \label{tab:ops}
\end{table}

\subsection{Operators}\label{sec:treams:ops}

The main idea underlying the class Operators and its subclasses is transferring the concept of abstract operators and their concrete representations from the mathematical domain to an implementation in Python. For example, if we want to apply a rotation by $\tfrac{\pi}{3}$ around the z-axis to a matrix $\mathbf{M}$, we can write this as
\begin{align}\label{eq:operator:example}
    \mathbf{R}\left(\tfrac{\pi}{3}\right) \mathbf{M} \mathbf{R}^{-1}\left(\tfrac{\pi}{3}\right)\,,
\end{align}
with $\mathbf{R}(\tfrac{\pi}{3})$ as an abstract rotation operator that depends on one variable. Depending on the basis in which we interpret the coefficients of $\mathbf{M}$, this abstract operator has different representations. If it is in the SW basis, these are Wigner-D matrix elements. With \treams, it is possible to write this as
\begin{lstlisting}[caption=Minimal example for the use of \texttt{Operator},label=lst:op]
>>> import numpy as np
>>> import treams
>>> class Foo:
>>>     def __init__(self, arr, basis):
...         self.arr = arr
...         self.basis = basis
...     def __array__(self):
...         return self.arr
...
>>> swb = treams.SphericalWaveBasis.default(1)
>>> mat = Foo(np.eye(6), swb)
>>> rot = treams.Rotate(np.pi)
>>> rot @ mat @ rot.inv
PhysicsArray(
...
\end{lstlisting}
where we first import \emph{numpy} and \treams and define a custom class with a minimal interface. Then, we create the matrix and the rotation operator, and, finally, the last command is an almost one-to-one correspondence of \cref{eq:operator:example}. \treams{} is only invoked three times, first to create a basis, second to create the rotation operator, and implicitly when using the @-symbol. Most of the lines are used to set up a custom class; of course, \treams{} comes with more elaborate implementations, but to emphasize the fact that the operator only depends on the presence of the attribute \texttt{basis}, we use the custom class here. This approach is chosen to facilitate custom class implementations by users if they need special functions and, thereby, improve the reusability of different parts of \treams.

There are several other operators implemented, which are listed in \cref{tab:ops}. Besides rotations, there are translations along a vector $\bm r$, the expansion in another basis which is separated in expansions with and without periodic boundaries. The operator \texttt{Expand} includes the expansion within the same basis set but at different positions, i.e., $\mathbf{C}^{(1)}$ and $\mathbf{C}^{(3)}$, but also the expansions as defined in \cref{eq:ex:pwsw,eq:ex:pwcw,eq:ex:cwsw}. The expansions with periodic boundary conditions create the matrices $\mathbf{\tilde C}^{(3)}$ when used within the same type of basis or the expansions of periodic waves in other basis sets, as in \cref{eq:exl:swpw,eq:exl:swcw,eq:exl:cwpw}. Additionally, the change between helicity and parity modes and the permutation of the lattice axes are implemented. The primary purpose of the latter is conveniently switching between the PW expansions of the scattering at a 2D lattice SWs in the x-y-plane and the expansion of the scattered waves at a lattice of CWs in the x-z-plane. We also implement operators to evaluate the electric, magnetic, or Riemann-Silberstein fields.

The operators can also be wrapped by the class \texttt{OperatorAttribute} and composed to a class to allow an alternative interface to the functionality by calling, e.g., in the example above, \texttt{mat.rotate(np.pi / 3)} in the last line of \cref{lst:op} to achieve the same result.

\subsection{Physics-aware arrays, T-matrices, and S-matrices}\label{sec:treams:arrays}

\begin{table}
    \centering
    \caption{Attributes of \texttt{PhysicsArray}}
    \begin{tabular}{lll}
        \toprule
         Description & Attribute & Type \\
         \midrule
         Vacuum wave number & \texttt{k0} & \texttt{float} \\
         Embedding material & \texttt{material} & \texttt{Material} \\
         Basis set & \texttt{basis} & \texttt{BasisSet} \\
         Polarization type & \texttt{poltype} & \texttt{str} \\
         Mode type (depending on basis set) & \texttt{modetype} & \texttt{str} \\
         Lattice & \texttt{lattice} & \texttt{Lattice} \\
         Wave Vector (along the lattice dimensions) & \texttt{kpar} & \texttt{WaveVector} \\
         \bottomrule
    \end{tabular}
    \label{tab:attrs}
\end{table}

In this section, we will motivate and highlight some use cases of the class \texttt{PhysicsArray}. The \texttt{PhysicsArray} objects behave much like \emph{numpy} arrays and can, when necessary, always be converted to them. However, they keep track of a set of relevant parameters attached to them during the computation. This set, also listed in \cref{tab:attrs}, currently contains the wave number, the material, the basis set, the polarization type, the mode type, and, for periodic structures, the underlying lattice and wave vector defining the phase relation between lattice sites. With polarization type, we refer to the helicity and parity basis. The mode type is relevant for the CWs and SWs, where it can be either regular or singular, and for the PWs if they are defined by two components to distinguish the direction $\uparrow$ or $\downarrow$ for the remaining third component. A set of these is attached to each dimension of the \texttt{PhysicsArray} and then compared during computations: most binary operations, such as addition or elementwise multiplications, compare the same dimensions; special operations, most importantly matrix multiplications, compare the last dimension of the first operand with the first dimension of the last operand. In simple cases, a single value can be attached to all dimensions, as shown here
\begin{lstlisting}[caption=Preservation of general attributes in \texttt{PhysicsArray},label=lst:pa:simple]
>>> import treams
>>> treams.PhysicsArray([[0, 1], [2, 3]], k0=1.2)
PhysicsArray(
    [[0, 1],
     [2, 3]],
    k0=1.2,
)
>>> treams.PhysicsArray([[0, 1], [2, 3]], k0=1.2) + [4, 5]
PhysicsArray(
    [[ 4,  6],
     [ 6, 8]],
    k0=1.2,
)
\end{lstlisting}
where the wave number is kept track of throughout the computation. However, it's also possible to attach one value per dimension. Here, two different wave numbers are attached to the two dimensions of the array:
\begin{lstlisting}[caption=Preservation of attributes in \texttt{PhysicsArray} per dimension,label=lst:pa:complex]
>>> import treams
>>> treams.PhysicsArray([[0, 1], [2, 3]], k0=(1.2, 3.4))
PhysicsArray(
    [[0, 1],
     [2, 3]],
    k0=(1.2, 3.4),
)
>>> treams.PhysicsArray([[0, 1], [2, 3]], k0=(1.2, 3.4)) + [4, 5]
PhysicsArray(
    [[ 4,  6],
     [ 6, 8]],
    k0=(1.2, 3.4),
)
>>> treams.PhysicsArray([[0, 1], [2, 3]], k0=(1.2, 3.4)) @ [4, 5]
PhysicsArray(
    [ 5, 23],
    k0=1.2,
)
>>> treams.PhysicsArray([0, 1], k0=1.2) + treams.PhysicsArray([2, 3], k0=3.4)
AnnotationWarning: at index 0: overwriting key 'k0'
PhysicsArray(
    [2, 4],
    k0=1.2,
)
\end{lstlisting}
It can be seen that the addition keeps both values in the result. For the matrix-vector multiplication, the result is a vector where the wave number corresponds to the value of the first matrix dimension.
The last command shows that if the parameters are not the same, a warning is issued to notify the user of potential errors.

These parameters are also exposed as attributes of the class. In total, they are sufficient to get a concrete representation of all operators described in the previous section. So, although the implementation of the class \texttt{Operator} and the associated \texttt{OperatorAttribute} is agnostic to the object which provides the parameters to get a concrete representation, the class \texttt{PhysicsArray} obviously is well suited to work with the implemented operators seamlessly.

Turning to the T-matrices now, we can build up on the class \texttt{PhysicsArray}, which already provides many useful operations. In fact, the class \texttt{TMatrix} derives from that class with additional restrictions on the parameters, for example, the basis set must be a \texttt{SphericalWaveBasis}. Also, it is supplemented with a range of additional methods to construct the T-matrix, e.g., from a single sphere by \texttt{TMatrix.sphere} or from a cluster of particles, and to evaluate its properties, e.g., the scattering cross-section or duality breaking.
In the very short example, we create the incident plane wave and the T-matrix of a core-shell particle, where the core is chiral.
\begin{lstlisting}[caption=T-matrix of a core-shell particle and,label=lst:tmat]
>>> import numpy as np
>>> import treams
>>> inc = treams.plane_wave([np.sin(np.pi / 4), 0, np.cos(np.pi / 4)], [0, 1, 0])
>>> mat_core = treams.Material(epsilon=4, kappa=0.1)
>>> mat_shell = treams.Material(epsilon=2 + 0.1j)
>>> tm = treams.TMatrix.sphere(
    lmax=6,
    k0=2 * np.pi,
    radii=[.4, .5],
    materials=[mat_core, mat_shell, 1]
)
>>> sca = (tm @ treams.Expand(basis=inc.basis).inv) @ inc
\end{lstlisting}
The scattered field is then calculated in the last line by
\begin{equation}
    \bm p_{\text{vsw}}
    = (\mathbf{T}_{\text{vsw}} \mathbf{E}_{\text{vpw},\text{vsw}}^{-1}) \bm a_{\text{vpw}}\,,
\end{equation}
where we added subscripts to indicate the type of basis. The matrix $\mathbf{E}_{\text{vpw},\text{vsw}}^{-1}$ contains the expansion coefficients of \cref{eq:ex:pwsw}. Note that the SW basis or the wave number is not explicitly specified but taken from the T-matrix object \texttt{tm}. If we continued in the example, we could now evaluate the electric field at specific points with \texttt{sca.efield}.
Similar to the T-matrices for SWs, the class \texttt{TMatrixC} implements functions for T-matrices in the \texttt{CylindricalWaveBasis}. The T-matrix can be solved analytically for cylinders, possibly with multiple concentric cylindrical shells, which is implemented in \treams as \texttt{TMatrixC.cylinder}. The primary usage of the class is similar to the one shown in \cref{lst:tmat}.

The S-matrix in the PW basis has a somewhat special role. A single \texttt{SMatrix} holds the coefficients for incoming waves propagating in one particular direction scattered into outgoing waves of one direction. However, as seen in \cref{eq:smatrix}, a full description needs four S-matrices that cover all combinations of directions. Thus, an instance of the class \texttt{SMatrices} holds four instances of \texttt{SMatrix} with matching parameters.

All the different classes shown here have many more methods implemented. However, we refer to the online documentation (\url{https://tfp-photonics.github.io/treams}) for detailed information on all functionalities. Especially, it also includes a range of simple examples with full scripts, as listed in \cref{tab:examples}. These examples are automatically rerun with each new version of \treams{} to ensure they stay up-to-date.

\section{Examples}\label{sec:examples}

\begin{table}
    \centering
    \caption{Covered examples in the online documentation.}
    \begin{tabular}{lll}
        \toprule
         Section & Example name & Description\\
         \midrule
         T-matrices & Spheres & T-matrix of spheres and their cross-section \\
         & Cluster & Global and local T-matrices of a finite cluster of particles \\
         & One-dimensional arrays & Chains of particles and their interaction \\
         & Two-dimensional arrays & Metasurface of spheres and their interaction \\
         & Three-dimensional arrays & Mode calculation in a crystal \\
         Cylindrical T-matrices & Cylinders & Cross-width of infinitely long cylinders \\
         & Transition from T-matrices & Cylindrical T-matrix from a chain of spheres \\
         & Cluster & Interaction of multiple cylindrical T-matrices \\
         & One-dimensional arrays & A one-dimensional grating of cylindrical T-matrices \\
         & Two-dimensional arrays & Eigenmodes in a two-dimensional array \\
         S-Matrices & Slabs & Transmittance and reflectance of a chiral slab \\
        & From T-matrices & Convert an array of T-matrices to an S-matrix \\
         & From cylindrical T-matrices & Convert an array of cylindrical T-matrices to an S-matrix \\
         & Band structure & Band structure calculation in periodic media \\
         \bottomrule
    \end{tabular}
    \label{tab:examples}
\end{table}

To demonstrate some capabilities of \treams, we present in the following a range of examples in the following. We start by demonstrating the different lattice sums and how they complement each other. Then, we show a more practical example of finding bound states in the continuum in metasurfaces.

\begin{figure}
    \centering
    \begin{minipage}[t]{.03\linewidth}\vspace{0pt}%
        \textsf{(a)}
    \end{minipage}%
    \begin{minipage}[t]{.45\linewidth}\vspace{0pt}%
        \includegraphics[width=\linewidth]{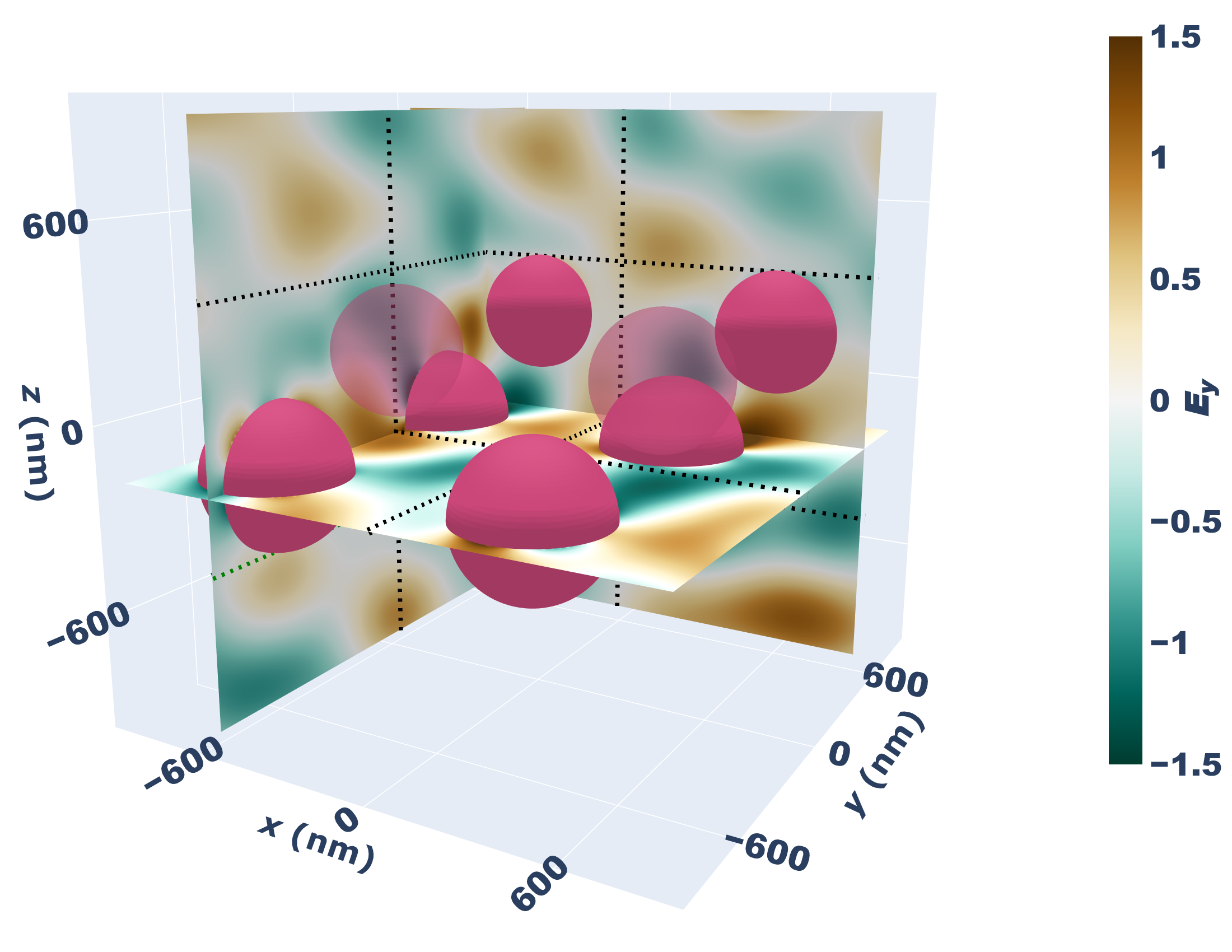}
    \end{minipage}%
    \begin{minipage}[t]{.03\linewidth}\vspace{0pt}%
        \textsf{(b)}
    \end{minipage}%
    \begin{minipage}[t]{.45\linewidth}\vspace{0pt}%
        \includegraphics[width=\linewidth]{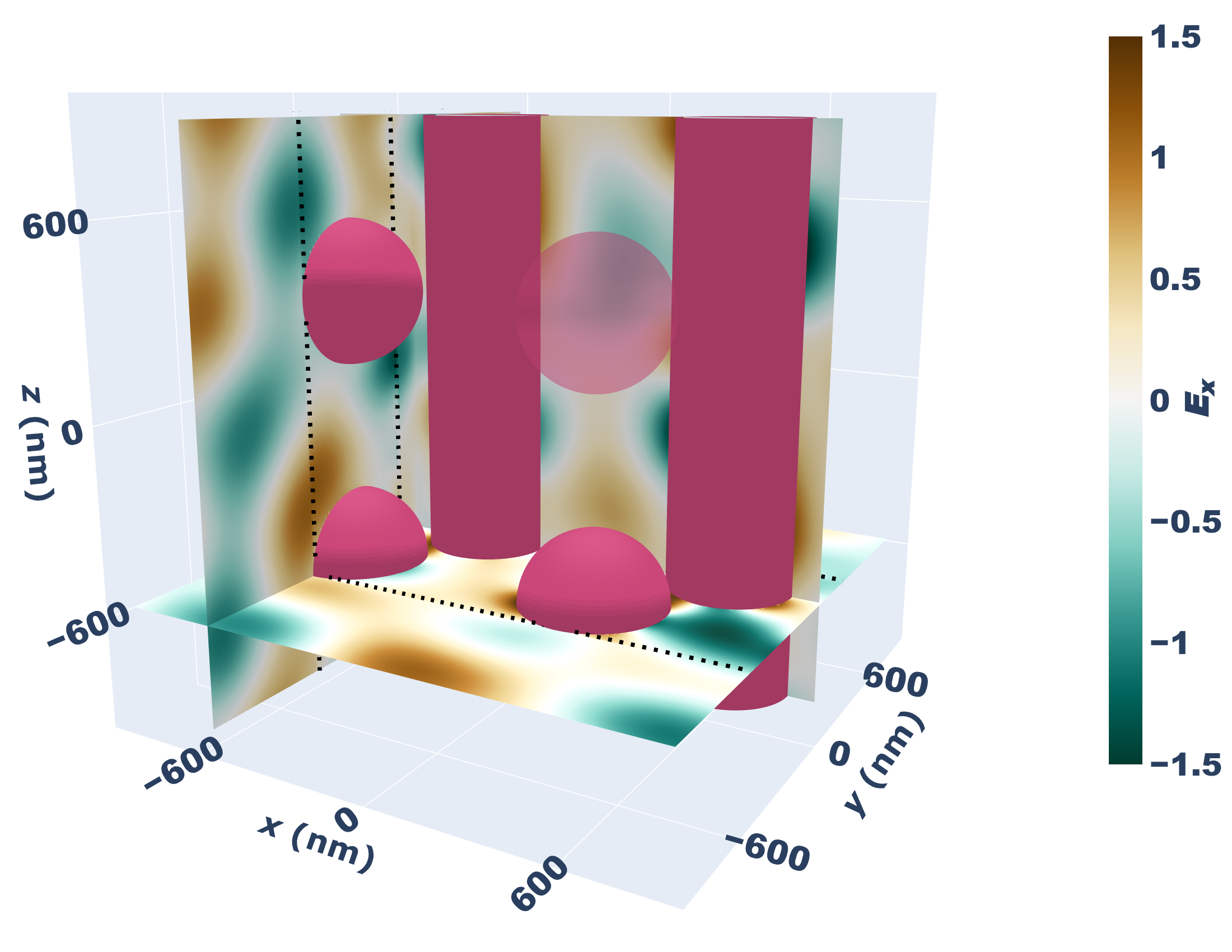}
    \end{minipage}
    \caption{The electric field in two different periodic arrangements.
    The scatterers are shown in red, and one electric field component is shown in three cross-sections through the 3D domain. Scatterers outside of these cross-sections are shown semi-transparent.\\
    Panel (a) shows four unit cells of a 2D lattice of spheres in the x-y-plane. The unit cell consists of two spheres with radii \SI{75}{\nano\meter} and \SI{65}{\nano\meter}. The separation between the spheres in the z-direction is \SI{120}{\nano\meter}. Therefore, they cannot be coupled by using an intermediate PW expansion. They must be treated as a whole with a SW expansion in a lattice using a complex unit cell. To create the electric field plot, different domains of validity are spliced. The fields above and below the bounding planes for the whole unit cell are computed in the PW expansion. The fields between these planes -- the region between the spheres -- have to be computed by calculating lattice sums of SWs. The repeated calculation of these lattice sums, with and without shifts perpendicular to the lattice plane, requires efficient implementations.\\
    Panel (b) shows another example of the electric field calculation. Four unit cells of a 2D lattice in the x-z-plane are shown. This calculation is done by combining all three basis sets. First, the coupling within 1D chains of spheres along the z-axis is calculated, and the result is expressed as a cylindrical T-matrix. Then, a complex unit cell of the sphere chain and the cylinder is formed, and their coupling is calculated for the periodic arrangement in the x-direction. Due to the different regions of validity, this step-wise procedure is necessary. The electric field is now spliced from PW expansions outside the bounding rectangular region. We can use the CW expansion between the solid cylinder and the circumscribing cylinder for the chain of spheres. Between the spheres themselves, only the SW expansion is valid.\\
    The absence of sharp features at the bounding boxes of the different domains indicates that the methods complement each other well when using sufficiently many expansion orders.
    }
    \label{fig:fields}
\end{figure}

\subsection{Complex unit cells in 1D and 2D lattices}\label{sec:examples:fields}

This section contains two examples: a 2D array of a unit cell consisting of two different spheres and a combination of spheres and cylinders in another array. These systems are shown in \cref{fig:fields}. Each panel shows one electric field component in three cross-sections through the domain.

In panel (a), the simulated system consists of gold spheres with radii \SI{225}{\nano\meter} and \SI{195}{\nano\meter} in an array with \SI{900}{\nano\meter} lattice pitch. They are positioned in the unit cell with a relative shift of $\bm r = (\SI{240}{\nano\meter}, \SI{300}{\nano\meter}, \SI{360}{\nano\meter})$ from the center of the bigger to the center of the smaller sphere. The lattice is illuminated with a TE polarized plane wave and oblique incidence at the wavelength \SI{600}{\nano\meter}. One electric field component is shown in the sections through the computational domain. The shown values are spliced into different parts. Between the spheres, only the SW expansion is valid, and lattice sums translating the scattered fields from all particles to each probed point are necessary. This requires highly efficient summation methods. The region above and below the array are computed with PWs. The absence of visible discontinuities, where the splicing occurs, shows how well these different expansions fit together. Another line, where the expansions are spliced, is at the boundary between the four shown elementary unit cells. Furthermore, there, these domains are seamlessly spliced.

Panel (b) shows a slightly more complicated system that includes SWs, CWs, and PWs. It consists of gold spheres and cylinders that are shifted by $\bm r = (\SI{300}{\nano\meter}, \SI{360}{\nano\meter}, \SI{240}{\nano\meter})$ to each other. First, the spheres are coupled together as a chain along the z-axis and transformed from a SW basis to CWs. Then a complex unit cell containing the chain and additionally the cylinder is formed using the CWs. Again, we splice four unit cells and also different parts depending on the convergence domains of the different sums. Above and below the spheres, only the SW expansion is valid. Between the cylinders and the spheres, the CW expansion is used. The PW expansion is used for smaller and larger values of $y$ according to \cref{fig:validity}(c).

\subsection{Quasi-bound states in the continuum}\label{sec:examples:bic}

\begin{figure}
    \centering
    \includegraphics[width=.95\linewidth]{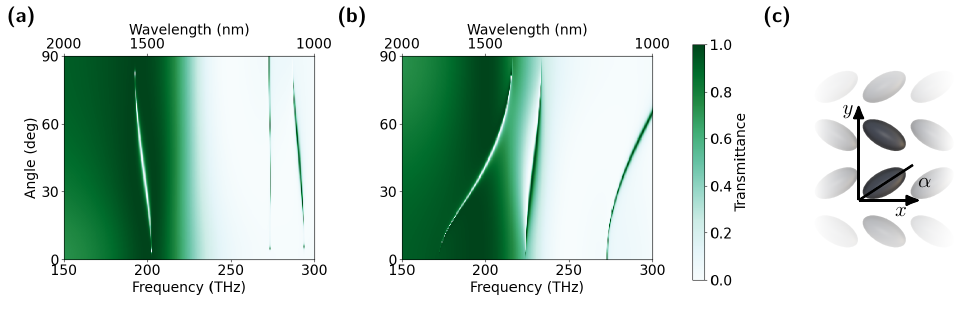}
    \caption{Quasi-BICs in a two-dimensional array of ellipsoids. Each unit cell contains two ellipsoids, which are rotated by an angle $\alpha$ to the x-axis. When the angle $\alpha$ is not zero or ninety degrees, the BIC in the lattice becomes a quasi-BIC and is thus visible in the transmission spectrum. Panels (a) and (b) show the result for normal incidence and linearly polarized light along the x- or y-axis, respectively. Panel (c) shows a unit cell of the lattice.}
    \label{fig:qbic}
\end{figure}

This example is concerned with finding quasi-BICs by introducing asymmetry in two-dimensional arrangements of particles. The complete script to reproduce the data using \treams{} is provided in \ref{app:qbic:code}. The T-matrix is provided in the supplementary materials together with all necessary files to reproduce it. The unit cell of the lattice is rectangular with side length $a_x = \SI{500}{\nano\meter}$ and $a_y = \SI{1000}{\nano\meter}$ and contains two identical particles, each described by a T-matrix of maximal multipole order 5. The T-matrix is computed using the software JCMsuite~\cite{burger2008a} and its built-in functionality to evaluate the multipole decomposition of the scattered field in a finite element method simulation. The particles are ellipsoids with short radius $2^{-\frac{1}{4}}\times\SI{210}{\nano\meter}$ and long radius $2^{\frac{1}{4}}\times\SI{210}{\nano\meter}$. Thus, the aspect ratio of these prolate ellipsoids is $\sqrt{2}$. The relative permittivity is set to 12.25, mimicking silicon in the wavelength range between \SI{1000}{\nano\meter} and \SI{2000}{\nano\meter}. The unit cell is shown in \cref{fig:qbic}(c) together with the coordinate axes and the definition of the angle $\alpha$ for the asymmetry. The panels (a) and (b) of \cref{fig:qbic} show the transmittance for linearly x- and y-polarized light. In the case of zero and ninety degrees, the BICs are not visible due to the symmetry of the lattice, but with increasing angle $\alpha$, they become quasi-BICs and appear as fine lines in the transmission spectrum. At \SI{1500}{\nano\meter} and $38^\circ$ angle, the two quasi-BIC lines for the two polarizations, shown in panels (a) and (b), are crossing.

\section{Conclusion\label{sec:conclusion}}

We presented the program \treams{} that provides a comprehensive framework to perform scattering calculations for a wide range of scenarios. Among the main features are the possibilities to perform T-matrix calculation in periodic systems with complex unit cells, i.e., unit cells with multiple objects described in a local basis, and the inclusion of chiral material parameters. The required lattice sums are implemented using exponentially fast converging series. Besides the case of periodic boundaries, standard T-matrix calculations for finite systems can also be done with typical operations such as rotations and translations included. Besides T-matrices using SWs, also CWs are implemented.
Scattering in stratified media is implemented using an S-matrix description for PWs.
The functionality is provided in two ways: a low-level interface that provides maximum speed and flexibility and a high-level interface that provides more convenient access to these functions and additional plausibility and error checks.
Furthermore, the T-matrices can be stored and loaded in an HDF5 file format.
The package \treams{} is open-source with permissive licensing. Moreover, it is accompanied by an extensive online documentation, including many examples and a large set of tests. By using continuous integration, the documentation, doctests included in the documentation, and tests are automatically run. Building pre-compiled packages for all three major operating systems, Windows, MacOS, and Linux, is also part of the automation pipeline and are available from the official Python Package Index.

There are multiple future avenues for the development of \treams. Regarding speed, making use of parallelization and GPU computation are apparent improvements. Adding automatic differentiation would enhance the capabilities of the program regarding possible optimization schemes.

We showed several examples where \treams{} can be applied: solving multi-scattering problems with arbitrary periodic boundary conditions with complex unit cells and finding quasi-BICs by breaking symmetries in a lattice. Furthermore, the methods described and implemented were used in various other publications. Those include the design of cavities for enhanced sensing of chiral molecules~\cite{feis2020a,scott2020,beutel2021a}, multi-scale simulation of molecular arrays in cavities~\cite{zerulla2022,zerulla2023}, T-matrix based homogenization~\cite{zerulla2023a}, the analysis of flat bands in moir\'{e} structures~\cite{dams2023a} and analytical models for metasurfaces~\cite{rahimzadegan2022}.

\section*{Acknowledgments}

D.B. and C.R. acknowledge support by the Deutsche Forschungsgemeinschaft (German Research Foundation) under Germany’s Excellence Strategy via the Excellence Cluster 3D Matter Made to Order (Grant No. EXC – 2082/1 – 390761711) and from the Carl Zeiss Foundation via CZF-Focus@HEiKA. The authors are grateful to the company JCMwave for their free provision of the FEM Maxwell solver JCMsuite. The authors would like to thank Stefan Mühlig, Martin Fruhnert, and Achim Groner that developed earlier versions of similar codes. 

\bibliographystyle{elsarticle-num}
\bibliography{treams}

\appendix

\section{Cylindrical and Spherical Coordinates}\label{app:coords}

We use the convention
\begin{subequations}
    \begin{align}
        x &= \rho \cos\phi = r \sin\theta \cos\phi \\
        y &= \rho \sin\phi = r \sin\theta \sin\phi \\
        z &= r \cos\theta
    \end{align}
\end{subequations}
for the definition of the cylindrical coordinates $\rho$, $\phi$, and $z$ and of the spherical coordinates $r$, $\theta$, and $\phi$. Note that the definition of $\phi$ coincides in these coordinate sets. The unit vectors are defined by
\begin{subequations}
    \begin{align}
        \erho &= \cos\phi \ex + \sin\phi \ey \\
        \er &= \sin\theta \cos\phi \ex + \sin\theta \sin\phi \ey + \cos\theta \ez \\
        \etheta &= \cos\theta \cos\phi \ex + \cos\theta \sin\phi \ey - \sin\theta \ez \\
        \ephi &= -\sin\phi \ex + \cos\phi \ey\,.
    \end{align}
\end{subequations}
If these coordinates and unit vectors are used for wave vector $\bm k$ instead of the real space coordinates $\bm r$, we use it as a subscript, for example, $\ephi_{\bm k}$. The exception is the radial vector $\bm{\hat k}$.

\section{Spherical harmonics and associated Legendre polynomials}\label{app:sphharm}
We define the spherical harmonics by
\begin{align}
    Y_{lm}(\theta, \phi)
    = \sqrt{\frac{2l + 1}{4\pi}\frac{(l - m)!}{(l + m)!}}
    P_l^m(\cos\theta)
    \ee^{\ii m \phi}\,,
\end{align}
where the associated Legendre polynomials are given by
\begin{align}
	P_l^m (x)
	=
	\frac{(-1)^m}{2^l l!}
	\left(1 - x^2\right)^{\frac{m}{2}}
	\frac{\dd^{l + m}}{\dd x^{l + m}}
	(x^2 - 1)^l
\end{align}
for $l \in \mathbb N_0$ and $m \in \{-l, -l + 1, \dots, l\}$.

\section{Translation coefficients for vector spherical and cylindrical waves}\label{app:transl}
The translation coefficients for SWs~\cite{cruzan1962,stein1961,tsang1985}
\begin{subequations}
    \begin{align}
    A_{l'm' lm}^{(n)}(\bm{r}, k) &=
    (-1)^m
    \ii^{l'-l}
    \sqrt{\pi\frac{2l'+1}{l'(l'+1)}\frac{2l+1}{l(l+1)}}
    \notag \\ \cdot
    \sum_{p}
    &\ii^p \sqrt{2p+1}
    z_p^{(n)}(kr) Y_{p,m-m'}(\theta, \phi)
    \begin{pmatrix}
    l & l' & p \\
    m & -m' & -m+m'
    \end{pmatrix}
    \begin{pmatrix}
    l & l' & p \\
    0 & 0 & 0
    \end{pmatrix}
    \left[l(l+1)+l'(l'+1) - p(p+1)\right]
    \\
    B_{l'm' lm}^{(n)}(k\bm{r}) &=
    (-1)^m
    \ii^{l'-l}
    \sqrt{\pi\frac{2l'+1}{l'(l'+1)}\frac{2l+1}{l(l+1)}} \notag \\ \cdot
    \sum_{p}
    &
    \ii^p \sqrt{2p+1}z_p^{(n)}(kr) Y_{p,m-m'}(\theta, \phi)
    \begin{pmatrix}
    l & l' & p \\
    m & -m' & -m+m'
    \end{pmatrix}
    \begin{pmatrix}
    l & l' & p-1 \\
    0 & 0 & 0
    \end{pmatrix}
    \sqrt{\left[(l+l'+1)^2-p^2\right]\left[p^2-(l-l')^2\right]}
    \,,
    \end{align}
\end{subequations}
where the sum index $p$ runs over all values where the Wigner 3j-symbols are non-zero,
are used to expand fields in the same type, namely incident in incident fields and scattered in scattered fields, using
\begin{align}
    \bm A_{lm,\pm}^{(n)}(\bm r, k)
    &=
    \sum_{l = 1}^\infty \sum_{m = -l}^l
    \left(
        A_{l'm',lm}^{(1)}(\bm r - \bm r', k)
        \pm B_{l'm',lm}^{(1)}(\bm r - \bm r', k)
    \right)
    \bm A_{l'm',\pm}^{(n)} (\bm r', k)
\end{align}
and to expand scattered fields in incident fields by
\begin{align}
    \bm A_{lm,\pm}^{(3)}(\bm r, k)
    &=
    \sum_{l = 1}^\infty \sum_{m = -l}^l
    \left(
        A_{l'm',lm}^{(3)}(\bm r - \bm r', k)
        \pm B_{l'm',lm}^{(3)}(\bm r - \bm r', k)
    \right)
    \bm A_{l'm',\pm}^{(1)} (\bm r', k)
    \,.
\end{align}
With a similar structure but simpler expressions, we have
\begin{align}
    \bm A_{k_z m,\pm}^{(n)}(\bm r, k)
    &=
    \sum_{m = -\infty}^\infty
    J_{m - m'}\left(\sqrt{k^2 - k_z^2} \rho_{\bm r - \bm r'}\right)
    \ee^{\ii (m - m') \phi_{\bm r - \bm r'} + \ii k_z (z - z')}
    \bm A_{k_zm',\pm}^{(n)} (\bm r', k)
\end{align}
and
\begin{align}
    \bm A_{k_z m,\pm}^{(3)}(\bm r, k)
    &=
    \sum_{m = -\infty}^\infty
    H_{m - m'}^{(1)}\left(\sqrt{k^2 - k_z^2} \rho_{\bm r - \bm r'}\right)
    \ee^{\ii (m - m') \phi_{\bm r - \bm r'} + \ii k_z (z - z')}
    \bm A_{k_zm',\pm}^{(1)} (\bm r', k)
\end{align}
for CWs~\cite{huang2019}.

\section{Permutation of Cartesian axes}\label{app:perm}
Due to the different alignment of the arrays for the different basis sets, it is necessary to permute the labels of the Cartesian axes. As an example, the 1D lattice of CWs is in the x-z-plane, but the 2D lattice of SWs is conventionally placed in the x-y-plane. To couple such systems, it is necessary to permute the corresponding basis sets. Also, the definition of the S-matrices is mostly suited to x-y-periodicity. Therefore, we need the transformation from a coordinate system with $x$, $y$, and $z$ to a system with $x = y'$, $y = z'$, and $z = x'$. Thus, the wave vector is $\bm k = k_x \ex + k_y \ey  + k_z \ez = k_z \ex' + k_x \ey' + k_y \ez'$ in these systems. This change leaves the scalar product $\bm k \bm r$ and, so, the exponential of the PWs invariant. However, the polarization vectors $\bm m_{\bm k}$ and $\bm n_{\bm k}$ of the PWs $\bm M_{\bm k}$ and $\bm N_{\bm k}$ change their definition.

In the original system, we have
\begin{subequations}
    \begin{align}
        \bm m_{\bm k} &= \ii \frac{k_y \ex - k_x \ey}{k_{xy}} \\
        \bm n_{\bm k} &= \frac{-k_x k_z \ex - k_y k_z \ey + k_{xy}^2 \ez}{k k_{xy}}\,,
    \end{align}
\end{subequations}
with $k_{xy} = \sqrt{\smash[b]{k_x^2 + k_y^2}}$, whereas the same definition leads to
\begin{subequations}
    \begin{align}
        \bm m_{\bm k}'
        &= \ii \frac{k_x \ex' - k_z \ey'}{k_{xz}}
        = \ii \frac{k_x \ez - k_z \ex}{k_{xz}} \\
        \bm n_{\bm k}'
        &= \frac{-k_z k_y \ex' - k_x k_y \ey' + k_{xz}^2 \ez'}{k k_{xz}}
        = \frac{-k_z k_y \ez - k_x k_y \ex + k_{xz}^2 \ey}{k k_{xz}}
    \end{align}
\end{subequations}
in the new coordinate system with $k_{xz}$ defined analogous to $k_{xy}$. Now, we can calculate the projection of the polarizations $\bm m_{\bm k}$ and $\bm n_{\bm k}$ onto the new polarizations $\bm m_{\bm k}'$ and $\bm m_{\bm k}'$. Remembering the minus sign, when projecting onto $\bm m_{\bm k}'$ we obtain
\begin{subequations}
    \begin{align}
        -\bm m_{\bm k}' \bm m_{\bm k}
        &= -\frac{k_y k_z}{k_{xy} k_{xz}} \\
        -\bm m_{\bm k}' \bm n_{\bm k}
        &= -\ii\frac{k_x k_{xy}^2 + k_x k_z^2}{k k_{xy} k_{xz}}
        = -\ii\frac{k_x k}{k_{xy} k_{xz}} \\
        \bm n_{\bm k}' \bm m_{\bm k}
        &= -\ii\frac{k_x k_y^2 + k_x k_{xz}^2}{k k_{xy} k_{xz}}
        = -\ii\frac{k_x k}{k_{xy} k_{xz}} \\
        \bm n_{\bm k}' \bm n_{\bm k}
        &= \frac{-k_z k_y k_{xy}^2 + k_z k_y k_x^2 - k_z k_y k_{xz}^2}{k^2 k_{xy} k_{xz}}
        = -\frac{k_y k_z}{k_{xy} k_{xz}}
        \,.
    \end{align}
\end{subequations}
So, the expansion coefficients of the PWs in the original system get transformed by multiplying with the matrix
\begin{align}
    \frac{1}{k_{xy} k_{xz}}
    \begin{pmatrix}
        -k_y k_z & -\ii k k_x \\
        -\ii k k_x & -k_y k_z
    \end{pmatrix}\,.
\end{align}
In the special case that $k_x = k_y = 0$, the matrix becomes
\begin{align}
    \begin{pmatrix}
        0 & -\ii \\
        -\ii & 0
    \end{pmatrix}
\end{align}
and for $k_x = k_z = 0$ it is
\begin{align}
    \begin{pmatrix}
        -\frac{k_y}{k} & 0 \\
        0 & -\frac{k_y}{k}
    \end{pmatrix}
    \,.
\end{align}
In the helicity basis, the two helicities are not mixed, but the expansion coefficient values have to be adjusted by multiplying with $\tfrac{-k_y k_z \mp \ii k k_x}{k_{xy} k_{xz}}$ for helicity $\pm1$.
The matrix for the inverse transformation $x = z''$, $y = x''$, and $z = y''$
\begin{align}
    \frac{1}{k_{xy} k_{yz}}
    \begin{pmatrix}
        -k_x k_z & \ii k k_y \\
        \ii k k_y & -k_x k_z
    \end{pmatrix}
\end{align}
can be obtained by applying the forward transformation twice with a correct relabelling of the entries.

\section{Expansions of lattice sums in other basis sets}\label{app:expandlattice}
To find expansions of periodic arrangements of scattered CWs of SWs in other basis sets, we use the integral expressions~\cite{wittmann1988}
\begin{subequations}\label{eq:swint}
\begin{align}
    \bm M_{lm}^{(3)}(\bm r, k)
    &=
    \frac{1}{2\pi \ii^l}
    \iint_{\mathbb{R}^2} \frac{\dd k_x \dd k_y}{k^2 \gamma_{xy}}
    \bm X_{lm}(\theta_{\bm k}, \phi_{\bm k}) \ee^{\ii (k_x x + k_y y + k_z z)}
    =
    \frac{1}{2\pi \ii^l}
    \iint_{\mathbb{R}^2} \frac{\dd k_x \dd k_z}{k^2 \gamma_{xz}}
    \bm X_{lm}(\theta_{\bm k}, \phi_{\bm k}) \ee^{\ii (k_x x + k_y y + k_z z)}
    \\
    \bm N_{lm}^{(3)}(\bm r, k)
    &=
    \frac{1}{2\pi \ii^{l - 1}}
    \iint_{\mathbb{R}^2} \frac{\dd k_x \dd k_y}{k^2 \gamma_{xy}}
    \bm{\hat k} \times \bm X_{lm}(\theta_{\bm k}, \phi_{\bm k}) \ee^{\ii (k_x x + k_y y + k_z z)}
    =
    \frac{1}{2\pi \ii^{l - 1}}
    \iint_{\mathbb{R}^2} \frac{\dd k_x \dd k_z}{k^2 \gamma_{xz}}
    \bm{\hat k} \times \bm X_{lm}(\theta_{\bm k}, \phi_{\bm k}) \ee^{\ii (k_x x + k_y y + k_z z)}\,,
\end{align}
\end{subequations}
where $\gamma_{ij} = \sqrt{1 - \frac{k_i^2 + k_j^2}{k^2}}$ for the SWs. There is a branch cut in the expansions at the $z = 0$ plane for the first integral and $y = 0$ plane for the second integral representation. The unspecified component of the integrals, $k_z$ in the first one and $k_y$ in the second one, is defined such that the PW propagates or decays when moving away from the branch cut. Thus, it is $k_z = \pm k \gamma_{xy}$ for $z \gtrless 0$ in the first case and analogous for $k_y$ in the second case.
For the CWs, we use~\cite{han2007}
\begin{subequations}\label{eq:cwint}
\begin{align}
    \bm M_{k_z,m}^{(3)}(\bm r, k)
    &=
    \frac{1}{\pi \ii^{m+1}}
    \int_{-\infty}^\infty \frac{\dd k_x}{k \gamma_{xz}}
    \ephi_{\bm k} \ee^{\ii (k_x x + k_y y + k_z z + m \phi_{\bm k})}
    \\
    \bm N_{k_z,m}^{(3)}(\bm r, k)
    &=
    -\frac{1}{\pi \ii^m}
    \int_{-\infty}^\infty \frac{\dd k_x}{k \gamma_{xz}}
    \etheta_{\bm k} \ee^{\ii (k_x x + k_y y + k_z z + m \phi_{\bm k})}\,,
\end{align}
\end{subequations}
with $k_y = \pm k \gamma_{xz}$ for $z \gtrless 0$.
We will use Poisson's formula
\begin{align}\label{eq:poisson}
    \sum_{\bm R \in \Lambda_d} \ee^{\ii \bm k \bm R}
    =
    \frac{(2\pi)^d}{V_d}\sum_{\bm G \in \Lambda_d^\ast} \delta^{(d)}(\bm k - \bm G)
\end{align}
to evaluate lattice sums over these solutions. $V_d$ is the generalized d-dimensional volume of a unit cell; in one dimension, it is equal to the lattice pitch $a$, and in two dimensions, it corresponds to the unit cell area $A$. This transformation is useful since the resulting delta distributions simplify the integrals.

Starting with SWs on a 2D lattice in the x-y-plane, we thereby get
\begin{align}
    \sum_{\bm R \in \Lambda_2}
    \bm M_{lm}^{(3)}(\bm{r} - \bm{R}, k)
    \ee^{\ii \bm k_\parallel \bm R}
    &=
    \frac{2\pi N_{lm}}{A k^2 \ii^l}
    \sum_{\bm G \in \Lambda_2^\ast}
    \left(
        \ii \pi_{lm}(\theta_{\bm k}) \etheta_{\bm k}
        - \tau_{lm}(\theta_{\bm k}) \ephi_{\bm k}
    \right)
    \frac{
        \ee^{\ii (\bm k \bm r + m \phi_{\bm k})}
    }{
        \sqrt{1 - \frac{(\bm k_\parallel + \bm G)^2}{k^2}}
    }
\end{align}
from combining \cref{eq:swint,eq:poisson}, where $\bm k = \bm k_\parallel + \bm G \pm \sqrt{\smash[b]{k^2 - (\bm k_\parallel + \bm G)^2}} \ez$ for $z \gtrless 0$ on the right-hand side. We can directly identify the PWs now. Similarly, we can derive the relations for the TM modes. In total, the result can be written as
\begin{align}
    \sum_{\bm R \in \Lambda_2}
    \begin{pmatrix}
        \bm M_{lm}^{(3)}(\bm{r} - \bm{R}, k) \\
        \bm N_{lm}^{(3)}(\bm{r} - \bm{R}, k)
    \end{pmatrix}
    \ee^{\ii \bm k_\parallel \bm R}
    &=
    -\frac{2\pi\ii N_{lm}}{A k^2 \ii^l}
    \sum_{\bm G \in \Lambda_2^\ast}
    \frac{
        \ee^{\ii m \phi_{\bm k}}
    }{
        \sqrt{1 - \frac{(\bm k_\parallel + \bm G)^2}{k^2}}
    }
    \begin{pmatrix}
        \tau_{lm}(\theta_{\bm k}) & \pi_{lm}(\theta_{\bm k}) \\
        \pi_{lm}(\theta_{\bm k}) & \tau_{lm}(\theta_{\bm k})
    \end{pmatrix}
    \begin{pmatrix}
        \bm M_{\bm k_\parallel + \bm G, \sign(z)}(\bm{r}, k) \\
        \bm N_{\bm k_\parallel + \bm G, \sign(z)}(\bm{r}, k)
    \end{pmatrix}
\end{align}
and after a change to helicity basis we get \cref{eq:exl:swpw}.

For SWs on a 1D lattice along the z-axis, we use the other form of the integral representation and get
\begin{align}
    \sum_{\bm R \in \Lambda_1}
    \bm M_{lm}^{(3)}(\bm{r} - \bm{R}, k)
    \ee^{\ii \bm k_\parallel \bm R}
    =
    \frac{N_{lm}}{a k \ii^l}
    \sum_{\bm G \in \Lambda_1^\ast}
    \int_{-\infty}^\infty
    \frac{\dd k_x \ee^{\ii (\bm k \bm r + m \phi_{\bm k})}}{k \sqrt{1 - \frac{k_x^2 + (k_\parallel + G)^2}{k^2}}}
    \left(
        \ii \pi_{lm}(\theta_{\bm k}) \etheta_{\bm k}
        - \tau_{lm}(\theta_{\bm k}) \ephi_{\bm k}
    \right)\,,
\end{align}
where we used that $\bm G = G \ez$ and $\bm k_\parallel = k_\parallel \ez$. The wave vector is $\bm k = k_x \ex \pm \sqrt{\smash[b]{k^2 - k_x^2 - (k_\parallel + G)^2}} + (k_\parallel + G) \ez$ for $y \gtrless 0$. Here, we can identify the integral representation of the CWs in \cref{eq:cwint}. Again, an equivalent procedure for the TM mode leads to the total result
\begin{align}
    \sum_{\bm R \in \Lambda_1}
    \begin{pmatrix}
        \bm M_{lm}^{(3)}(\bm{r} - \bm{R}, k) \\
        \bm N_{lm}^{(3)}(\bm{r} - \bm{R}, k)
    \end{pmatrix}
    \ee^{\ii \bm k_\parallel \bm R}
    =
    -\frac{\ii \pi N_{lm}}{a k \ii^{l - m}}
    \sum_{\bm G \in \Lambda_1^\ast}
    \begin{pmatrix}
        \tau_{lm}(\theta_{\bm k}) & \pi_{lm}(\theta_{\bm k}) \\
        \pi_{lm}(\theta_{\bm k}) & \tau_{lm}(\theta_{\bm k})
    \end{pmatrix}
    \begin{pmatrix}
        \bm M_{k_\parallel + G, m}^{(3)}(\bm{r}, k) \\
        \bm N_{k_\parallel + G, m}^{(3)}(\bm{r}, k)
    \end{pmatrix}\,,
\end{align}
which can then be converted to \cref{eq:exl:swcw}.

Finally, the CWs on a 1D lattice along the x-axis can be rewritten as
\begin{align}
    \sum_{\bm R \in \Lambda_1}
    \bm M_{k_z, m}(\bm r - \bm R, k) \ee^{\ii \bm k_\parallel \bm R}
    &=
    \frac{2}{ak \ii^{m + 1}}
    \sum_{\bm G \in \Lambda_1^\ast}
    \frac{\ee^{\ii (\bm k \bm r + m \phi_{\bm k})}}{\sqrt{1 - \frac{(k_\parallel + G)^2 + k_z^2}{k^2}}}
    \ephi_{\bm k}\,,
\end{align}
where we use $\bm k_\parallel = k_\parallel \ex$, $\bm G = G \ex$, and $\bm k = (k_\parallel + G) \ex \pm \sqrt{k^2 - (k_\parallel + G)^2 - k_z^2} \ey + k_z \ez$ for $y \gtrless 0$ and in total we obtain
\begin{align}
    \sum_{\bm R \in \Lambda_1}
    \begin{pmatrix}
        \bm M_{k_z, m}^{(3)}(\bm r - \bm R, k) \\
        \bm N_{k_z, m}^{(3)}(\bm r - \bm R, k)
    \end{pmatrix}
    \ee^{\ii \bm k_\parallel \bm R}
    &=
    \frac{2}{ak \ii^m}
    \sum_{\bm G \in \Lambda_1^\ast}
    \frac{1}{\sqrt{1 - \frac{(k_\parallel + G)^2 + k_z^2}{k^2}}}
    \begin{pmatrix}
        1 & 0 \\ 0 & 1
    \end{pmatrix}
    \begin{pmatrix}
        \bm M_{\bm{\hat{k}}}(\bm r, k) \\
        \bm N_{\bm{\hat{k}}}(\bm r, k)
    \end{pmatrix}\,,
\end{align}
which is \cref{eq:exl:cwpw} in helicity basis.

\section{Script for the quasi-BIC example}
\label{app:qbic:code}

The following script can be used to recalculate the data for \cref{fig:qbic}

\begin{lstlisting}[caption=Commented quasi-BIC calculation,label=lst:qbic]
import pickle

import joblib

import numpy as np
import treams
import treams.io


def run_xy(tm, lat, positions, theta, phi):
    """Calculate transmittance for x- and y-polarization.

    For two objects described by the same T-matrix, calculate the transmittance of x-
    and y-polarized plane waves, when the objects are arranged in a lattice and have
    the given positions in their unit cell. The objects are rotated by theta and then
    plus and minus phi, respectively.

    Args:
        tm (treams.TMatrix): T-matrix.
        lat (Sequence): Lattice definition.
        positions (Sequence): Positions in the unit cell.
        theta (float): Rotation about the y-axis.
        phi (float): Rotation about the z-axis.

    Returns:
        Tuple
    """
    # Create lattice from definition
    lattice = treams.Lattice(lat)

    # Build cluster T-matrix
    cluster = treams.TMatrix.cluster(
        [tm.rotate(phi, theta, 0), tm.rotate(-phi, theta, 0)], np.array(positions)
    )
    # Calculate the interaction in the lattice for normal incidence
    kpar = [0, 0]
    cluster = cluster.latticeinteraction.solve(lattice, kpar)
    pwb = treams.PlaneWaveBasisByComp.default([0, 0])
    pwx = treams.plane_wave(
        kpar,
        [1, 0, 0],
        basis=pwb,
        k0=tm.k0,
        material=1,
        poltype="parity",
        modetype="up",
    )
    pwy = treams.plane_wave(
        kpar,
        [0, 1, 0],
        basis=pwb,
        k0=tm.k0,
        material=1,
        poltype="parity",
        modetype="up",
    )
    # Calculate S-matrix from the array
    sm = treams.SMatrices.from_array(cluster, pwb)
    return sm.tr(pwx)[0], sm.tr(pwy)[0]


def wrapper(tm):
    """Take a T-matrix and run a sweep over angles."""
    theta = np.pi / 2
    angles = np.linspace(0, np.pi / 2, 200)
    lattice = [500, 1000]
    positions = [[0, -250, 0], [0, 250, 0]]
    return [run_xy(tm, lattice, positions, theta=theta, phi=phi) for phi in angles]


tms = treams.io.load_hdf5("ellipsoid.h5")
# We use joblib for parallel execution
res = joblib.Parallel(n_jobs=8)([joblib.delayed(wrapper)(tm) for tm in tms])

# Store in pickle file
with open("qbics.pickle", "w") as fobj:
    pickle.dump(res, fobj)

\end{lstlisting}

The necessary T-matrix file in the HDF format is available in the supplementary materials. Furthermore, we also provide scripts and the necessary files to recalculate the T-matrix with JCMsuite or COMSOL in the supplementary materials. For JCMsuite, the necessary configuration files are used by calling \texttt{ellipsoid\_jcm.py}, which generates the HDF5 file directly. For the COMSOL calculation, the java file needs to be compiled with \texttt{comsol compile ellipsoid\_comsol.java} and then run with \texttt{comsol batch ellipsoid\_comsol.class}. The resulting coefficients are then exported as a text file, that can be converted to the suitable T-matrix format with the script \texttt{ellipsoid\_comsol.py}. The calculation with COMSOL returns the T-matrix in the helicity format, such that the listing above needs to be changed accordingly in lines 45 and 54.

\end{document}